\RequirePackage{fix-cm}
\documentclass[smallextended,envcountsect]{svjour3}
\usepackage{graphicx}
\usepackage{natbib}
\usepackage{url}
\usepackage{amsmath,amssymb,amsfonts}
\usepackage{booktabs}
\usepackage{listings}
\usepackage{multirow}
\usepackage{makecell}
\usepackage{xspace}
\usepackage{soul}
\usepackage{tabularx}
\usepackage{tcolorbox}
\tcbuselibrary{skins}
\usepackage{textcomp}
\usepackage{colortbl}
\usepackage{algorithm}
\usepackage[hidelinks]{hyperref}
\usepackage{caption}
\usepackage{todonotes}
\usepackage{longtable}
\usepackage{array}
\usepackage{float}
\usepackage{placeins}
\usepackage[T1]{fontenc}
\usepackage[utf8]{inputenc}

\newcolumntype{L}[1]{>{\raggedright\arraybackslash}p{#1}}
 
\newcommand{\myprompt}[2]{%
  \begin{tcolorbox}[
    enhanced,
    colback=gray!10,
    colframe=black!20,
    boxrule=0.4pt,
    arc=1mm,
    left=2mm,right=2mm,top=1mm,bottom=1mm,
    fonttitle=\bfseries,
    coltitle=black,
    title={#1},
    before skip=3pt plus 2pt minus 1pt,
    after skip=3pt plus 2pt minus 1pt
  ]
  \small #2
  \end{tcolorbox}%
}

\newcommand{\rqconclusion}[2]{%
  \begin{tcolorbox}[
    enhanced,
    colback=black!4,
    colframe=black!55,
    boxrule=0.6pt,
    arc=1mm,
    left=2mm,right=2mm,top=1.5mm,bottom=1.5mm,
    fonttitle=\bfseries,
    coltitle=white,
    title={#1},
    before skip=6pt plus 2pt minus 1pt,
    after skip=8pt plus 2pt minus 1pt
  ]
  \small #2
  \end{tcolorbox}%
}

\newif\ifDEBUG
\DEBUGfalse

\newif\ifEXTENDED
\EXTENDEDfalse

\usepackage{algorithm}
\usepackage{xspace}
\usepackage{multirow} %
\usepackage{fancyvrb} %
\usepackage{pifont}

\usepackage{enumitem}
\setlist[itemize]{leftmargin=*,noitemsep,topsep=0pt}
\setlist[enumerate]{leftmargin=*}

\makeatletter
\patchcmd{\@makecaption}
	{\scshape}
	{}
	{}
	{}
\makeatletter
\patchcmd{\@makecaption}
	{\\}
	{.\ }
	{}
	{}
\makeatother
\def\tablename{Table}

\newcommand{\eg}{\textit{e.g.,}\xspace}
\newcommand{\etal}{\textit{et al.}\xspace}

\DeclareMathSymbol{\mlq}{\mathord}{operators}{``}
\DeclareMathSymbol{\mrq}{\mathord}{operators}{`'}

\newif\ifSAVESPACE
\SAVESPACEfalse

\ifSAVESPACE

\else

\fi

\usepackage{soul}
\usepackage{fontawesome}

\ifDEBUG
    \newcommand{\AH}[1]{\todo[color=cyan,inline]{AH:#1}}
    \newcommand{\AM}[1]{\todo[color=red,inline]{Machiry:#1}}
    \newcommand{\JD}[1]{\todo[color=yellow,inline]{JD:#1}}
    \newcommand{\SA}[1]{\todo[color=green,inline]{SA:#1}}
    \newcommand{\PA}[1]{\todo[color=orange,inline]{PA:#1}}
    
    \newcommand{\KR}[1]{\todo[color=yellow,inline]{Kyle:#1}}
    \newcommand{\LS}[1]{\todo[color=green,inline]{LS:#1}}
    \newcommand{\HP}[1]{\todo[color=cyan,inline]{HP:#1}}
    \newcommand{\NJE}[1]{\todo[color=red,inline]{NJE: #1}}
    \newcommand{\GKT}[1]{\todo[color=red,inline]{GKT:#1}}
    
    \newcommand{\RH}[1]{\todo[color=red,inline]{RH:#1}}
    \newcommand{\WJ}[1]{\todo[color=SkyBlue,inline]{Wenxin:#1}} 
    \newcommand{\KC}[1]{\todo[color=orange,inline]{Kelechi Says:#1}}
    \newcommand{\AG}[1]{\todo[color=orange,inline]{AG:#1}}
    \newcommand{\PJ}[1]{\todo[color=lime,inline]{PJ:#1}}
    \newcommand{\AZ}[1]{\todo[color=teal,inline]{Antonio:#1}}
    
    \newcommand{\PT}[1]{\todo[color=pink,inline]{Parth:#1}}

\else
    \newcommand{\AH}[1]{}
    \newcommand{\AM}[1]{}
    \newcommand{\JD}[1]{}
    \newcommand{\SA}[1]{}
    \newcommand{\PA}[1]{}
    \newcommand{\KR}[1]{}
    \newcommand{\LS}[1]{}
    \newcommand{\HP}[1]{}
    \newcommand{\NJE}[1]{}
    \newcommand{\GKT}[1]{}
    \newcommand{\KC}[1]{}
    \newcommand{\RH}[1]{}
    \newcommand{\WJ}[1]{}
    \newcommand{\AG}[1]{}
    \newcommand{\PJ}[1]{}
    \newcommand{\PT}[1]{}
    \newcommand{\AZ}[1]{}
    
\fi

\usepackage{cleveref}
\crefformat{section}{\S#2#1#3}
\crefname{figure}{Figure}{Figures}
\crefname{table}{Table}{Tables}
\crefname{theorem}{Theorem}{Theorems}
\crefname{thm}{Theorem}{Theorems}
\crefname{lemma}{Lemma}{Lemmata}
\crefname{equation}{Eqt.}{Eqts.}
\crefformat{Grammar}{Grammar #1}
\crefname{appendix}{Appendix}{Appendices}
\crefname{listing}{Listing}{Listings}

\newcommand{\myparagraph}[1]{\paragraph{#1}}
\renewcommand{\myparagraph}[1]{\vspace{0.25em} \noindent \underline{\textit{#1:}}}
\usepackage{url}

\makeatletter
\newcommand{\linebreakand}{%
  \end{@IEEEauthorhalign}
  \hfill\mbox{}\par
  \mbox{}\hfill\begin{@IEEEauthorhalign}
}
\makeatother

\usepackage{pifont}

 \usepackage{pifont}%

\journalname{Preprint}
\renewcommand{\makeheadbox}{}

\begin{document}

\title{How Do Coding Agents Optimize Software and Report Performance Validation? A Large-Scale Empirical Study of Open-Source Pull Requests}
\titlerunning{How Do Coding Agents Optimize Software and Report Performance Validation?}

\author{Huiyun Peng$^{\star}$ \and Ricardo Calvo$^{\star}$ \and Kelechi G. Kalu \and James C. Davis}
\authorrunning{H. Peng et al.}

\institute{H. Peng \at
              Purdue University, West Lafayette, Indiana, USA \\
              \email{peng397@purdue.edu}
           \and
           R. Calvo \at
              Purdue University, West Lafayette, Indiana, USA\\
              \email{rcalvome@purdue.edu}
           \and
           K. G. Kalu \at
              Purdue University, West Lafayette, Indiana, USA\\
              \email{kalu@purdue.edu}
           \and
           J. C. Davis \at
              Purdue University, West Lafayette, Indiana, USA\\
              \email{davisjam@purdue.edu}
          }

\date{}

\maketitle
\begingroup
\renewcommand{\thefootnote}{}
\footnotetext{$^{\star}$These authors contributed equally to this work.}
\endgroup

\begin{abstract}
Software performance optimization requires identifying bottlenecks and improvement opportunities, and empirically evaluating the effects and trade-offs across performance objectives.
Autonomous AI coding agents submit performance-oriented pull requests (PRs) to open-source projects, yet it remains unclear whether these changes reflect established optimization practices and adequately substantiate their claimed value.

Extending our MSR~2026 Mining Challenge pilot study of 407 performance PRs, we analyze a sample of 1,130 agentic and 1,130 human-authored performance PRs collected through June 2026.
We compare adoption outcomes and patch characteristics between the two groups, and examine their optimization practices and validation behaviors.
The results show that agentic PRs are merged less often than human-authored PRs (54.4\% vs.\ 73.3\%), but the two groups use similar optimization strategies and report validation at comparable rates. Agentic PRs rely more on static reasoning and less on benchmarks (46.1\% vs.\ 57.2\%), although benchmark use converges by the end of the study period. Agentic PRs also report validation for code-smell refactorings as often as for resource-targeting changes, whereas human-authored PRs do so less often.
Across both groups, about half of validated PRs report no quantitative performance metric, and relevant trade-offs are rarely quantified.
Agentic optimization practice has become increasingly similar to human practice, but the supporting evidence remains incomplete. Establishing whether agent-proposed optimizations are supported by rigorous, quantitative, and multidimensional evidence remains an important challenge. These findings motivate evaluation infrastructure and review criteria that measure intended effects and relevant costs rather than treating the presence of validation alone as sufficient.

\end{abstract}

\section{Introduction}
Software performance directly affects the reliability~\citep{Ulrich2016}, user experience~\citep{liu2014effects}, and energy efficiency~\citep{pereira2017energy} of modern systems.
Despite sustained advances in compiler optimization~\citep{muchnick1997advanced}, algorithm design~\citep{cormen1990introduction}, and hardware-aware tuning~\citep{bryant2003computer}, performance engineering remains labor-intensive and requires specialized expertise and systematic measurement~\citep{gregg2014systems}.
The difficulty arises in part because an optimization can have non-local effects, depend on assumptions about workloads and hardware, and require empirical evidence, such as benchmarking or profiling, to establish its impact~\citep{mytkowicz2009producing}.

Large language models (LLMs) increasingly support software development tasks such as code synthesis, debugging, and refactoring~\citep{hou2024large}, and have shown potential to improve non-functional properties such as execution time and energy efficiency~\citep{shypula2023learning, peng2026sysllmatic}.
More recently, autonomous coding agents have begun authoring substantial numbers of pull requests (PRs) to open-source projects~\citep{watanabe2026use, li2025rise}, including changes intended to improve performance.
However, empirical evidence about how agents optimize software in real-world projects remains limited.
Assessing agentic performance PRs requires examining what changes they make and what evidence supports their claimed benefits. 
A plausible optimization may lack measurements under representative workloads or overlook costs in other resource dimensions.
Examining the changes alongside their supporting evidence helps characterize agentic performance work and supports maintainers in evaluating agent-authored PRs.

This article extends our MSR~2026 Mining Challenge paper~\citep{peng2026agents}, which reported a pilot study of 407 performance-related PRs (324 agentic and 83 human-authored) drawn from AIDev~\citep{li2025rise}.
The pilot established the study's initial empirical framework: it adapted the SysLLMatic optimization catalog~\citep{peng2026sysllmatic} to real-world PRs, expanding it from 43 to 59 patterns, and developed a taxonomy for classifying reported validation evidence.
It also provided initial evidence that agentic and human-authored PRs used similar optimization strategies but differed in how frequently they reported validation.
The pilot comparison was limited, however, by its small, imbalanced sample and by different repository-inclusion thresholds for agentic and human-authored PRs ($\geq$100 versus $\geq$500 stars).

This work uses the pilot study as its methodological foundation and extends the investigation.
We reconstruct the AIDev collection pipeline, extending its coverage through June 2026---ten months beyond AIDev-v3---and apply the same repository criteria to both groups, producing a sample of 1{,}130 agentic and 1{,}130 human-authored performance PRs.
We retain the pilot's methodology for analyzing optimization patterns and validation practices as RQ2 and RQ3, formalize and expand its methodology for analyzing adoption and maintainability as RQ1, and introduce RQ4 to examine the quantitative metrics reported as validation evidence.
We ask:
\begin{itemize}
    \item \textbf{RQ1:} How do agentic and human-authored performance PRs differ in adoption outcomes, patch size, and structural indicators of maintainability?
    \item \textbf{RQ2:} Which optimization patterns occur in agentic and human-authored performance PRs, and how does their prevalence differ between the two groups?
    \item \textbf{RQ3:} How do agentic and human-authored performance PRs differ in their reported testing and validation practices, and how do these practices change over time?
    \item \textbf{RQ4:} What quantitative performance metrics, if any, are reported in the validation evidence of agentic and human-authored performance PRs, and how do those reported metrics relate to the applied optimization?
\end{itemize}

Our results suggest that agents have largely converged with humans in the optimizations they propose, but not in how they substantiate their effectiveness.
Agentic and human-authored PRs draw on much the same optimization strategies and report validation at comparable rates (82.3\% vs.\ 80.6\%), yet agentic PRs are merged less often (54.4\% vs.\ 73.3\%) and lean more on static reasoning than on benchmarks (46.1\% vs.\ 57.2\%).
Comparable rates also mask a difference in judgment: human-authored PRs report validation evidence for code-smell refactorings less often than for changes that target a performance resource (73.5\% vs.\ 82.8\%), whereas agentic PRs show no such distinction (83.0\% vs.\ 81.7\%).
More importantly, neither group routinely measures whether a change improves performance or what it costs: among PRs with validation evidence, only 50.9\% of agentic PRs and 52.4\% of human-authored PRs report a performance metric at all.
An important open problem for agent-authored optimization is therefore producing measurements that establish the intended effect and expose its trade-offs.

\vspace{1em}
\noindent
\textbf{\underline{Significance.}}
These findings provide a large-scale empirical baseline for evaluating this transition beyond patch production: they identify quantitative, multidimensional evaluation as a concrete capability gap for agent and tool designers; support review policies that require evidence that is representative and independently checkable, not plausible explanations alone; and establish measures for tracking whether future agents improve not only the optimizations they propose but also how convincingly they substantiate them.

\section{Background \& Related Work}
\label{sec:background}
This section provides the background for the four research questions and positions our study relative to prior work.
Section~\ref{sec:background-perf} introduces the performance-engineering concepts underlying our analysis: the optimization-pattern catalog used to characterize performance PRs (RQ2), forms of validation evidence (RQ3), and performance metrics (RQ4).
Section~\ref{sec:background-agentic} discusses the transition from AI-assisted to autonomous software engineering and LLM-based code optimization.
Section~\ref{sec:related-work} reviews related work on empirical studies of AI-generated performance changes.

\subsection{Performance Engineering}
\label{sec:background-perf}
\subsubsection{Optimization Patterns}
\label{sec:background-pattern}
Performance engineering spans the full software stack, from application logic to operating system and hardware behavior.
At the source-code level, performance tuning includes, for example, replacing an inefficient algorithm or data structure, caching repeated computation, improving memory locality, simplifying control flow, and parallelizing independent work.
These techniques are extensively studied across the algorithm, compiler, computer-architecture, and systems-performance literature~\citep{cormen1990introduction, muchnick1997advanced, bryant2003computer, gregg2014systems}, but the literature describes them at different levels of abstraction and for different purposes.

To answer RQ2, we require a concrete and consistently defined vocabulary for labeling the optimization strategies used in performance-related PRs. We adopt the SysLLMatic catalog~\citep{peng2026sysllmatic} because it consolidates 43 source-level optimization patterns into seven categories and demonstrates their use in guiding whole-program optimization across real-world benchmarks. Its organization at both category and pattern levels provides the granularity needed to compare broad strategy distributions while retaining the specific mechanism applied in each PR.

\subsubsection{Performance Validation}
\label{sec:background-validation}
Optimizations require rigorous validation because performance is highly sensitive to measurement noise, workload choice, and experimental design; best practices therefore emphasize representative workloads, repeatable protocols, and statistically grounded comparisons~\citep{hoefler2015scientific}. 
In software engineering practice, this evidence can take several forms.
First, \emph{Benchmark-Based Validation} relies on quantitative measurements (\eg unit tests or microbenchmarks) to directly assess performance impact~\citep{horky2015utilizing, hoefler2015scientific}.
Second, \emph{Profiling-Based Validation} uses profiling artifacts such as hotspot traces, CPU samples, or flame graphs to localize performance costs~\citep{gregg2016flame}.
Third, \emph{Static-Reasoning-Based Validation} supports performance claims through algorithmic arguments without relying on runtime data, \eg by identifying asymptotic inefficiencies~\citep{olivo2015static}. 
We use these three forms of validation evidence as the basis for answering RQ3, which examines how agentic and human-authored performance PRs report support for their performance claims.

\subsubsection{Performance Metrics and Trade-offs}
\label{sec:background-tradeoff}

Performance metrics characterize different dimensions of a system's behavior. 
The systems-performance literature commonly considers measures such as latency and execution time, throughput, memory footprint and allocation pressure, CPU work, I/O and network traffic, concurrency behavior, and cost~\citep{gregg2014systems}.
Other dimensions include artifact size, such as compiled code size~\citep{cummins2025llm}; build time, including compilation time~\citep{jayatilaka2021compile}; and energy consumption~\citep{pereira2017energy}.
Importantly, optimizations can affect these dimensions differently, either improving several metrics simultaneously or introducing trade-offs between them.
For example, eliminating redundant computation, dead code, or duplicated queries can reduce execution time and resource use simultaneously~\citep{muchnick1997advanced, sellis1988multiple}.
In contrast, caching and memoization can reduce execution time while increasing memory use~\citep{michie1968memo, gregg2014systems}; parallelization can reduce wall-clock time while increasing CPU and memory use~\citep{cormen1990introduction}; and compression can reduce bandwidth and storage requirements while increasing CPU work~\citep{abadi2006integrating}.
We summarize this systems-performance vocabulary in Table~\ref{tab:performance-dimensions} and use it as the basis for answering RQ4, which examines the quantitative performance metrics reported in validation evidence and how those metrics relate to the applied optimization.

\begin{table}[htbp]
\centering
\small
\caption{Performance dimensions and terminology drawn from the systems-performance literature and used as the conceptual basis for RQ4~\citep{gregg2014systems, jain1990art}. The identifiers M1--M9 link these concepts to the metric categories used in the subsequent analysis; the definitions and examples clarify the scope of each dimension.}
\label{tab:performance-dimensions}
\begin{tabularx}{\textwidth}{L{0.05\textwidth} L{0.24\textwidth} X}
\toprule
\textbf{ID} & \textbf{Dimension / term} & \textbf{Definition and examples} \\
\midrule
M1 & Latency / execution time & Elapsed time to respond to a request or complete a task. Wall-clock time includes waiting; tail latency (\eg p95 or p99) describes upper response-time percentiles. Typically reported in milliseconds or seconds. \\
M2 & Throughput & Amount of work completed per unit time, such as requests per second, operations per second, or tokens per second. \\
M3 & Memory & Memory occupied or allocated by a program. Footprint describes occupied memory, while allocation pressure concerns allocation activity and associated garbage-collection work. Examples include heap size, resident set size (RSS), and allocation counts. \\
M4 & CPU / compute work & Processor resources used to perform a task. CPU time measures processor time consumed; utilization measures the fraction of processor capacity in use. Other indicators include cycles, instructions, and function-call counts. \\
M5 & I/O and network & Communication and data-access activity, such as bytes transferred, database queries, network round trips, or API calls. Cache hit rate is the fraction of accesses served from a cache. \\
M6 & Artifact size & Storage or distribution size of a generated artifact, such as an executable binary, JavaScript bundle, or container image, typically measured in bytes. \\
M7 & Build and CI time & Elapsed time for development workflows, including compilation, incremental rebuilds, and continuous-integration (CI) test execution. \\
M8 & Energy and cost & Energy consumed and monetary expenditure associated with a workload. Energy is measured in joules or kilowatt-hours; power, measured in watts, is its rate of consumption. \\
M9 & Scalability / concurrency & How performance changes as workload or available resources grow, and how simultaneous work is handled. Examples include throughput under increasing load and latency with different worker or thread counts. \\
\bottomrule
\end{tabularx}
\end{table}

\subsection{Agentic Software Engineering}
\label{sec:background-agentic}

\subsubsection{From AI Assistance to Autonomous Software Engineering}
Early applications of code-trained large language models (LLMs) to software engineering primarily supported developer-directed tasks, such as code completion and local code synthesis~\citep{chen2021evaluating, nijkamp2022codegen, jiang2026survey}.
Their use has since expanded across the software development lifecycle, including code generation, program repair, testing, and refactoring~\citep{hou2024large, wei2023copiloting, wang2024software}.
In these AI-assisted settings, developers are responsible for integrating and evaluating generated changes.

More recent systems incorporate LLMs into autonomous coding agents that perform multi-step software-engineering tasks, including repository exploration, multi-file editing, tool invocation, and iterative revision based on tool feedback~\citep{yang2024swe, shinn2023reflexion, wang2025ai}.
Such agents have begun contributing pull requests (PRs) to open-source repositories, where their changes undergo the same review and integration processes as other contributions~\citep{li2025rise, watanabe2026use}.
This shift delegates to agents decisions that were previously made by developers, including how to assess a change and what evidence to provide in support of it~\citep{li2025rise, davis2026model}.
Understanding the reliability of these delegated decisions is therefore important for evaluating agent-authored software changes.

\subsubsection{LLM-Based Code Optimization}

A growing body of work uses LLMs to improve the performance of existing code~\citep{gong2025language}.
Some approaches learn optimization strategies from human-written code changes. PIE fine-tunes models on pairs of slow and fast competitive-programming submissions~\citep{shypula2023learning}, while Supersonic trains a sequence-to-sequence model on pairs of functionally equivalent C/C++ competitive-programming submissions to generate targeted, diff-based source-level optimizations~\citep{chen2024supersonic}.
Other approaches guide LLMs at inference time using execution feedback or iterative self-optimization, as in PerfCodeGen~\citep{peng2025perfcodegen} and EffiLearner~\citep{huang2024effilearner}. Related benchmarks, including EvalPerf and COFFE, evaluate the efficiency of code generated by LLMs~\citep{liu2024evaluating, peng2025coffe}.
LLMs have also been applied to lower layers of the software stack, including compiler optimization passes~\citep{cummins2025llm} and parallel and HPC programs through agentic systems~\citep{ wei2024improving}.
Energy has also been studied alongside execution time: LLMs can generate more energy-efficient code~\citep{peng2024large}, while subsequent work benchmarks the energy consumption of generated code~\citep{islam2025evaluating} and finds that runtime reductions can sometimes come with increased energy use~\citep{coignion2025faster}.
SysLLMatic~\citep{peng2026sysllmatic} extends this work to whole-program optimization by combining performance diagnostics and iterative refinement across languages, improving latency, throughput, energy, and memory.
Together, this body of work establishes that LLMs can generate effective code optimizations. Our study complements these efforts by examining how autonomous coding agents apply optimization strategies in real-world pull requests and how the accompanying validation evidence compares with that reported by human contributors.

\subsection{Related Work}
\label{sec:related-work}

Prior works have mined real-world patches to study performance optimization. RAPGen~\citep{garg2025rapgen} mines performance-related GitHub commits to build a transformation knowledge base for fixing code inefficiencies, demonstrating that historical patches encode reusable optimization actions. 
Yi \etal~\citep{yi2025ai} uses performance-improving commits with JMH benchmarks as ground truth, asks an LLM to optimize the pre-change code, and compares LLM-generated optimizations against both original and developer-authored improvements.
Concurrently, recent work has begun empirically examining agentic PRs in real-world repositories, comparing their characteristics and integration outcomes with human-authored PRs~\citep{watanabe2026use, li2025rise}.
Prior to our pilot study, no work had compared agentic and human-authored performance pull requests or analyzed performance factors such as optimization strategies and validation practices.
The pilot~\citep{peng2026agents} provided the first such comparison across 407 PRs and developed the optimization catalog and validation taxonomy used in this article.
This work applies those instruments to an extended sample of 2{,}260 PRs and broadens the analysis to temporal changes in validation and the quantitative metrics reported as validation evidence.

\begin{figure}[htbp]
    \centering
    \includegraphics[width=\linewidth, trim={0.7cm 2.5cm 2.3cm 0.3cm}, clip]{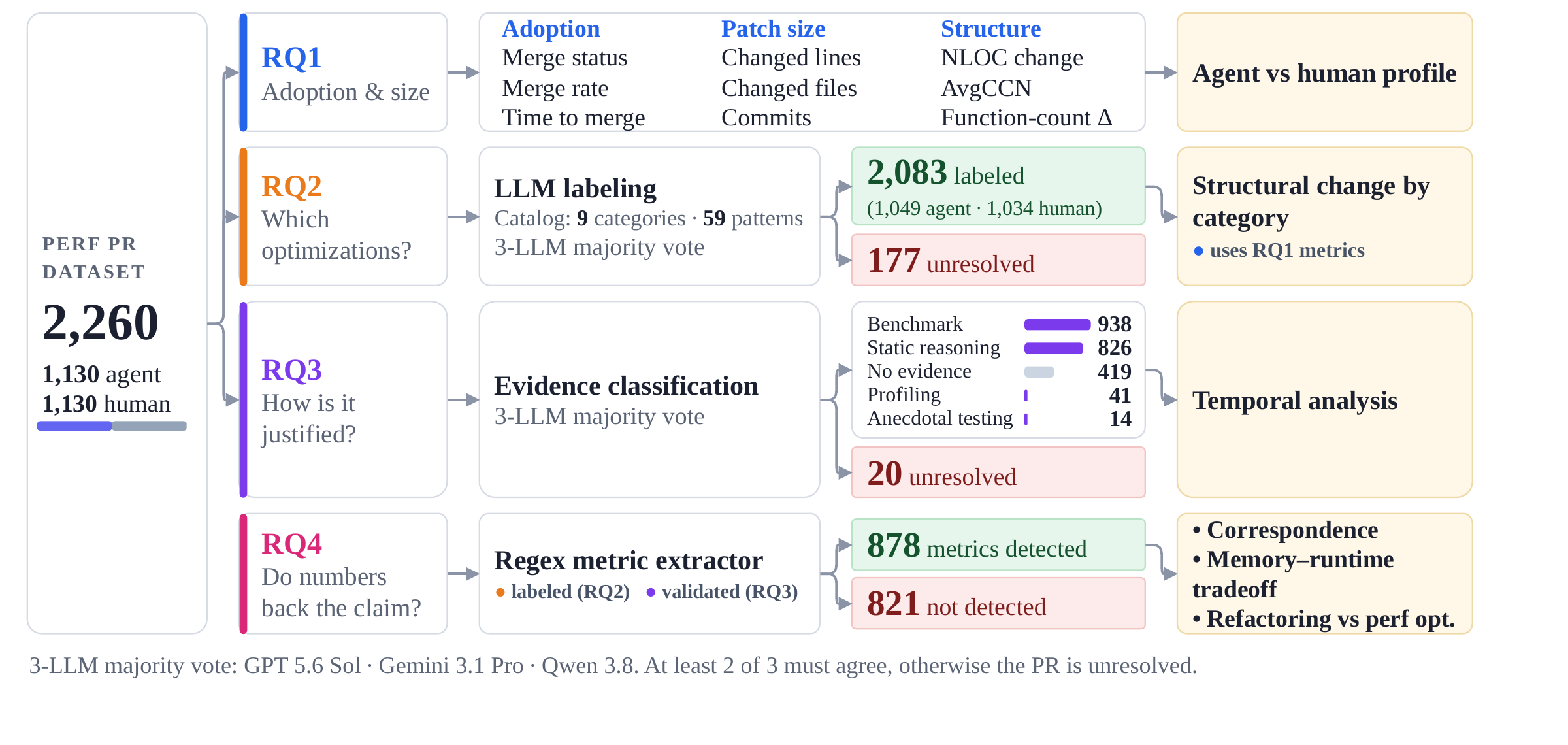}
    \caption{Overview of the analyses applied to the performance-PR sample. RQ1 measures adoption, patch size, and code structure; RQ2 assigns optimization patterns; RQ3 classifies reported validation evidence; and RQ4 extracts quantitative metrics from that evidence.}
    \label{fig:rq-analysis-overview}
\end{figure}

\section{Methodology}
This section describes dataset construction (Section~\ref{sec:data-prep}), the comparison of adoption outcomes, patch size, and structural indicators of maintainability for RQ1 (Section~\ref{sec:methodology-rq1}), the assignment of optimization-pattern labels for RQ2 (Section~\ref{sec:methodology-rq2}), the classification of validation evidence for RQ3 (Section~\ref{sec:methodology-rq3}), and the analysis of reported performance metrics for RQ4 (Section~\ref{sec:methodology-rq4}). Figure~\ref{fig:rq-analysis-overview} shows an overview of our methodology.

\begin{figure}[H]
    \centering
    \includegraphics[width=\linewidth, trim={1.2cm 17cm 1.2cm 4cm}, clip]{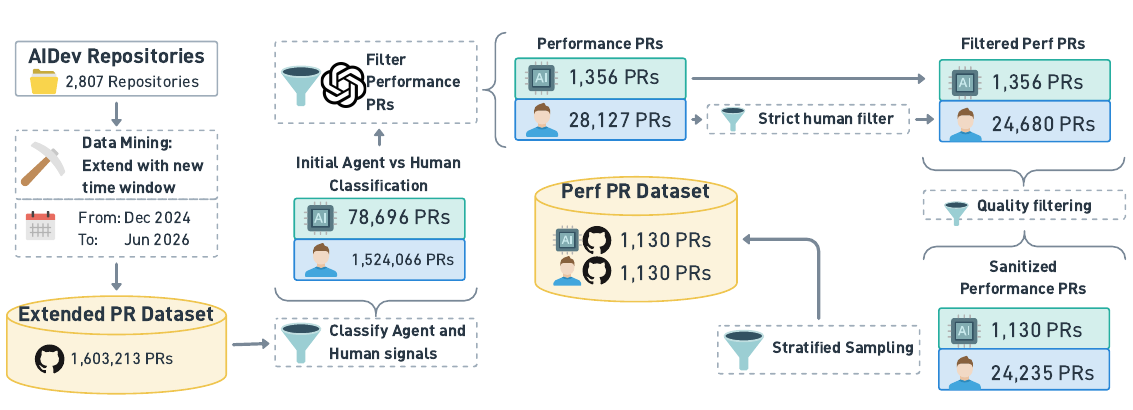}
    \caption{
        Dataset Construction.
        Starting with the repository frame defined by AIDev, we mine PRs submitted during the extended observation window, use AIDev's attribution signals to classify them as agentic or human-authored PRs, retain those classified as performance-related, apply the strict human-authored filter and quality criteria, and draw a 1:1 weekly stratified sample of agentic and human-authored performance PRs.
        Table~\ref{tab:current-sample-pipeline} reports the procedure and number of PRs retained at each stage.
    }
    \label{fig:methodology-overview}
\end{figure}

\subsection{Dataset Construction}
\label{sec:data-prep}
We first summarize the dataset used in the pilot study and the limitations that motivate its reconstruction (Section~\ref{sec:data-prep-prior}). We then describe the large-scale dataset-construction pipeline, including PR mining, authorship attribution, task-type classification, quality filtering, and balanced sampling (Section~\ref{sec:data-prep-current}). Figure~\ref{fig:methodology-overview} provides an overview of this pipeline.

\subsubsection{Pilot-Study Dataset}
\label{sec:data-prep-prior}
The pilot study was conducted using \textsc{AIDev}~\citep{li2025rise}, a dataset of GitHub pull requests (PRs) authored by agents and humans with commit, diff, review, and repository-level metadata. It used two subsets constructed under different repository-popularity thresholds: \textsc{AIDev-Pop}, containing 33{,}596 PRs from repositories with at least 100 stars, and \textsc{Human-PR}, containing 6{,}618 PRs from repositories with at least 500 stars. Restricting these subsets to PRs labeled \texttt{perf} yielded 340 agentic PRs and 88 human-authored PRs.

To support code-level analysis, the study retrieved fine-grained commit diffs from GitHub. It reconstructed 7{,}874 commit-level records for the 88 human-authored PRs; when combined with \textsc{AIDev-Pop}, the dataset contained 15{,}284 records from 428 performance PRs. 
Quality filtering removed records with missing filenames~(46), configuration-only changes~(44), merge commits~(10{,}166), and commits from deleted repositories~(74). 
The resulting dataset comprised 4{,}954 valid records from 407 PRs and retained coverage of 324/340 agentic PRs (95.3\%) and 83/88 human-authored PRs (94.3\%).

\myparagraph{Limitations of this dataset}
The initial dataset had three limitations for directly comparing agentic and human-authored performance PRs. First, the two groups were substantially imbalanced, with 324 agentic PRs and 83 human-authored PRs after filtering. Second, they were drawn using different repository-popularity thresholds, with agentic PRs collected from repositories with at least 100 stars and human-authored PRs from repositories with at least 500 stars. Third, the collection ended in August 2025 and therefore did not represent subsequent changes in agent activity and development practices. These limitations motivate the reconstructed dataset described next.

\subsubsection{Large-Scale Extension and Balanced Sample}
\label{sec:data-prep-current}
For the large-scale extension, we reconstruct and extend the AIDev dataset to address the limitations identified above. Specifically, we extend the observation period, draw agentic and human-authored PRs from the same repository population, and balance the two groups by submission week.
AIDev-v3's collection ends on August 1, 2025, \footnote{Our collection pipeline is based on \href{https://huggingface.co/datasets/hao-li/AIDev/tree/v3}{AIDev-v3} rather than the more recent AIDev-v4, which extends the collection through November 2025 and adds Google Jules as an additional agent.} limiting the comparison to an earlier period and missing more recent trends.
We therefore mine PR metadata from December 24, 2024 through June 1, 2026, extending the observation window by ten months and capturing more recent agentic activity.

\myparagraph{Mining and author-type attribution}
We reproduce AIDev's PR-classification pipeline over its frame of 2{,}807 repositories and apply the same threshold of at least 100 stars to both groups. Table~\ref{tab:current-sample-pipeline} summarizes each stage and identifies whether its procedure is derived from AIDev or introduced in the large-scale extension.
Repository-level mining yields 1{,}603{,}213 PRs and their associated commit, diff, review, and repository metadata.

We use AIDev's agentic and human-authored PR classification pipeline. This process classifies 78{,}696 PRs as \texttt{agentic} and 1{,}524{,}066 as \texttt{human\_authored}; the remaining 451 PRs are unresolved\footnote{451 PRs could not be classified as either agentic or human-authored.}.
Following AIDev, we define an agentic PR as one authored and submitted by an autonomous coding agent. Classification is based on confirmed signals, such as a known agent account or verified agent application, or probable signals, such as an agent-owned branch created during the agent's period of activity.\footnote{We reproduce AIDev's detection mechanisms. Confirmed signals are an exact agent account (\eg \texttt{devin-ai-integration[bot]}), a Copilot author paired with a \texttt{copilot/} branch, or an official GitHub App verified as an autonomous coding agent. Probable signals are a \texttt{codex/} or \texttt{cursor/} branch within the agent's valid activity period, an official agent task or session URL accompanied by implementation evidence, or an unambiguous ``implemented by [coding agent]'' statement. Signals of assistance alone are not agentic: a lone \texttt{Co-Authored-By} trailer, an AI badge, or a generated summary or review marks a PR \texttt{ai\_assisted}. A user author with no agentic or AI-assistance signal is a \texttt{human\_authored}.} A PR that exhibits only AI assistance---for example, model-generated code, review text, or a summary used under human control---is not classified as agentic.

\myparagraph{Task-type classification}
We follow AIDev's two-stage task-type classification procedure.
The first stage assigns labels from Conventional Commit title prefixes (\eg a title beginning with perf) without LLM inference;
PRs not classified by this heuristic proceed to the second stage, LLM classification.
The heuristic classifies 703{,}044 PRs, of which 7{,}952 are identified as performance-related.
We classify the remaining 900{,}169 PRs using AIDev's prompt with \texttt{gpt-5.6-luna} in place of AIDev's \texttt{gpt-4.1-mini}. 
This stage identifies an additional 21{,}531 performance-related PRs at a total inference cost of approximately \$110.
In total, the procedure identifies 29{,}483 performance-related PRs: 1{,}356 agentic PRs and 28{,}127 initial human-authored PRs.

We apply an additional filter to the human-authored PRs to reduce the risk of including automated or AI-assisted contributions. Specifically, the filter excludes PRs associated with a known coding-agent account, an author reported by GitHub as a bot, a suspicious automation-related user login, a coding-agent task or session URL, an AI coauthor trailer, an explicit AI-authorship statement, a structured AI-review signal, or AI-generated metadata. These rules exclude 3{,}447 of the 28{,}127 initial human-authored PRs, leaving 24{,}680 human-authored PRs.\footnote{The absence of these observable signals supports classification as human-authored but does not confirm human authorship. We therefore exclude AI-assisted PRs from both groups and compare agent-authored PRs with PRs showing no observable model involvement.}

\myparagraph{Accuracy of task-type classification}
Because our performance PR classifier uses a different LLM from AIDev, we first assess the agreement between the task-type labels produced by the two classification pipelines for PRs shared by both collections. This comparison does not treat either set of labels as ground truth. 
Using AIDev's labels as the reference for this pairwise comparison, our classifier obtains 64.4\% precision, 72.5\% recall, and an $F_1$ score of 68.2\%.

We separately construct a manually annotated ground-truth set to evaluate performance PR classification accuracy. 
We randomly sample 100 PRs from the mined dataset, and one author manually labels each PR as performance-related or non-performance-related based on its content. Against these annotations, our classifier achieves 83.3\% precision, 84.7\% recall, and an $F_1$ score of 84.0\%. 
To compare our classifier with the original AIDev classifier against the same ground truth, we further restrict the evaluation to the 28 annotated PRs that appear in both datasets. 
On this shared subset, our classifier achieves higher recall (82.4\% vs.\ 52.9\%) and $F_1$ score (80.0\% vs.\ 64.3\%), while AIDev achieves higher precision (81.8\% vs.\ 77.8\%).

\myparagraph{Perf PR quality filtering and sampling}
We next apply the same quality filters from the pilot study, excluding PRs with empty filenames, configuration-only changes, deleted repositories, or merge-only changes.
Among agentic PRs, the filters exclude 204 with empty filenames, 20 with configuration-only changes, and two with merge-only changes, leaving 1{,}130 of the initial 1{,}356 PRs.
Among human-authored PRs, they exclude 61, 375, and nine PRs for the same respective reasons, leaving 24{,}235 of the initial 24{,}680 PRs. In total, the quality filters exclude 671 PRs.

\begin{figure}[htbp]
    \centering
    \includegraphics[width=0.82\linewidth]{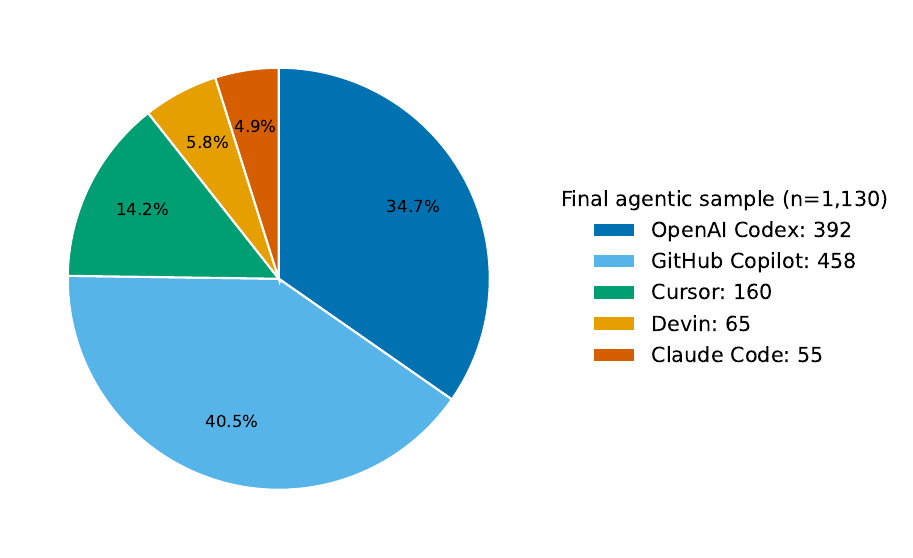}
    \caption{Distribution of the 1{,}130 agentic performance PRs in the final sample across the five agents covered by AIDev's attribution rules. Wedges show percentages within the agentic sample; the legend reports counts.}
    \label{fig:aidev-agent-distribution}
\end{figure}

To obtain a balanced sample, we retain all eligible agentic PRs and sample an equal number of human-authored PRs within each of the 65 ISO-week strata. The final analysis sample contains 2{,}260 PRs and is balanced 1:1 both overall and within each week. Figure~\ref{fig:aidev-agent-distribution} shows the distribution of the agentic sample across the five agents covered by AIDev's classification rules.

\myparagraph{Relationship to the pilot sample}
The large-scale sample was drawn from the full eligible population, rather than constructed by augmenting the pilot dataset. Of the pilot study’s 324 agentic PRs, 134 (41.4\%) appear in the large-scale sample; 45 are absent from the reconstructed collection, 75 no longer meet the performance classification, and 70 are excluded by the quality filters.\footnote{Quality-filter exclusions may reflect GitHub API errors, which occurred at a rate of 0.997\% during collection, or subsequent modifications to PR contents.}
None of the pilot study’s 83 human-authored PRs appear in the large-scale sample. Of these, 14 are absent from the reconstructed collection, 19 no longer meet the performance classification, 9 are excluded by the human-authored filter, and 4 are excluded by the quality filters. The remaining 37 were eligible but were not selected by the week-stratified random sampling procedure, which drew from 24{,}235 human-authored PRs without preference for pilot PRs. Overall, 134 of the pilot study’s 407 PRs (32.9\%) reappear in the large-scale sample. All results reported in this work are recomputed from the large-scale sample.

\begin{table}[htbp]
\centering
\small
\setlength{\tabcolsep}{4pt}
\caption{Dataset construction for the balanced analysis sample. ``AIDev-derived'' denotes a published AIDev procedure that we reproduced or adapted; extension-specific stages were introduced for the large-scale analysis.}
\label{tab:current-sample-pipeline}
\begin{tabularx}{\textwidth}{L{0.12\textwidth} L{0.18\textwidth} X L{0.20\textwidth}}
\toprule
\textbf{Stage} & \textbf{Origin} & \textbf{Procedure} & \textbf{Result} \\
\midrule
1. Mining & Extension-specific; AIDev repository frame & Mine all PRs in the common repository frame ($\geq 100$ stars) through June 1, 2026 & 1{,}603{,}213 PRs \\
2. Attribution & AIDev-derived; extension reconciliation & Apply five published attribution searches and preserve pinned historical positives & 78{,}696 agentic; 1{,}524{,}066 human-authored; 451 unresolved \\
3. Task type & AIDev-derived classifier; adapted model & Apply title-prefix rules, followed by AIDev's prompt with \texttt{gpt-5.6-luna}; retain performance PRs & 29{,}483 performance PRs; 1{,}573{,}724 non-performance PRs; 6 errors \\
4. human-authored filter & Extension-specific & Exclude observable bot, coding-agent, AI-authorship, AI-review, and generated-content signals & 28{,}127 $\rightarrow$ 24{,}680 human-authored PRs \\
5. Quality and sampling & Extension-specific & Apply quality criteria and deterministic 1:1 sampling within weeks & 1{,}130 agentic + 1{,}130 human-authored PRs \\
\bottomrule
\end{tabularx}
\end{table}

\subsection{Adoption, Patch Size, and Maintainability (RQ1)}
\label{sec:methodology-rq1}

The pilot study reported merge status, time to merge, and changes in NLOC, average cyclomatic complexity, and function count as descriptive characteristics of its sample~\citep{peng2026agents}.
RQ1 extends these measures to the large-scale dataset and compares agentic and human-authored PRs in their integration outcomes, patch sizes, and changes in code structure.

We measure adoption using PR merge status and, for merged PRs, the elapsed time between creation and merge. Merge rates are computed over all sampled PRs, whereas merge times are computed only for merged PRs.\footnote{Because merge status can change over time, these measures represent the state of each PR at the archived dataset snapshot rather than its final outcome.} To characterize patch size, we record added lines, deleted lines, changed files, and commits, and report each group's median.

As indicators of maintainability, we measure mean file non-comment lines of code (NLOC), mean file average cyclomatic complexity (AvgCCN~\citep{mccabe1976complexity}), and mean file function count. We compute these metrics with Lizard~\citep{yin2019lizard} for the base and head versions of every changed file, then calculate each PR's percentage change. Percentage changes allow comparison across PRs that modify files and codebases of different sizes. This analysis includes the 2{,}029 PRs for which both versions of every changed file could be retrieved (1{,}005 agentic and 1{,}024 human-authored). We exclude PRs with incomplete retrieval so that each before-after comparison covers the same set.

Merge status is categorical, so we use a chi-square test to determine whether merge rates differ by author type and report Cram\'er's $V$ as the effect size.
Merge time, patch-size measures, and structural changes are skewed, so we use Mann--Whitney $U$ tests to compare their distributions between agentic and human-authored PRs and report Cliff's $\delta$ as the effect size. 
Positive values of Cliff's $\delta$ indicate larger values for agentic PRs. 
To account for the multiple RQ1 comparisons, we apply the Benjamini--Hochberg correction to control the false discovery rate and report adjusted $q$-values.

\subsection{Optimization Patterns (RQ2)}
\label{sec:methodology-rq2}
To answer RQ2, we label each performance PR with an optimization pattern and compare the pattern distributions of agentic and human-authored PRs. 
We use the SysLLMatic catalog~\citep{peng2026sysllmatic}, introduced in Section~\ref{sec:background-pattern} and refined during the pilot study. 
Its categories describe broad strategies (\eg Memory and Data Locality Optimizations), while its patterns describe specific mechanisms (\eg Caching).
Section~\ref{sec:methodology-pattern-refinement} describes the catalog refinement conducted in the pilot study, Section~\ref{sec:methodology-rq2-labeling} presents the labeling procedure used in the large-scale extension, and Section~\ref{sec:methodology-rq2-analysis} specifies the category-level comparisons.

\subsubsection{Optimization Catalog}
\label{sec:methodology-pattern-refinement}
The SysLLMatic catalog was originally demonstrated on benchmark optimizations, including HumanEval~\citep{chen2021evaluating}, but had not been evaluated on real-world PRs, whose changes are larger and more heterogeneous.
The pilot study therefore refined the catalog through iterative model classification and human review.

As part of the refinement, two LLMs independently assigned an optimization category and pattern to each sampled PR using the original catalog. 
We then examined disagreements between their assignments to identify cases poorly represented by the catalog. 
Agreement was modest: Cohen's $\kappa$ was 0.53 for categories and 0.48 for patterns in agentic PRs, compared with 0.42 and 0.35 in human-authored PRs. Manual examination of the disagreements identified optimization strategies that were not represented in the original catalog.
Two authors, both Ph.D.\ students, independently assessed whether each case was adequately described by an existing pattern or required the addition, deletion, or revision of a pattern.
The first annotator proposed 15 additions, one deletion, and one revision, whereas the second proposed 10 additions, two deletions, and four revisions.
They subsequently compared their annotations, resolved disagreements through discussion, consolidated overlapping proposals, and revised the pattern definitions and category assignments.
This process continued until additional review produced no substantively new patterns, consistent with iterative category refinement in qualitative analysis~\citep{usman2017taxonomies}.
With the revised catalog, inter-model agreement increased to $\kappa=0.62$ at the category level and $\kappa=0.47$ at the pattern level for agentic PRs, and to $\kappa=0.61$ and $\kappa=0.47$, respectively, for human-authored PRs.

The refinement expanded the catalog from 43 to 59 patterns by adding 17 patterns, deleting 1, and revising 5.
It also introduced two high-level categories: \textit{Build, Compilation, and Infrastructure Optimization} and \textit{Network, Database, and Data Access Optimization}. Table~\ref{tab:new-patterns} presents the patterns assigned to these new categories, and Table~\ref{tab:pattern-catalog} summarizes the final catalog of 59 patterns across nine categories.

\begingroup
\setlength{\tabcolsep}{5pt}
\small
\begin{longtable}{L{0.20\textwidth} L{0.22\textwidth} L{0.50\textwidth}}
\caption{The nine patterns in the two high-level categories added during catalog refinement: five under \textit{Build, Compilation, and Infrastructure Optimization} and four under \textit{Network, Database, and Data Access Optimization}. Both categories capture optimizations that recur in real-world performance PRs but fall outside the source-level scope of the original SysLLMatic catalog.}
\label{tab:new-patterns} \\
\toprule
\textbf{Category} & \textbf{Pattern} & \textbf{Description} \\
\midrule
\endfirsthead
 
\multicolumn{3}{c}{\tablename\ \thetable{} -- continued from previous page} \\
\toprule
\textbf{Category} & \textbf{Pattern} & \textbf{Description} \\
\midrule
\endhead
 
\midrule
\multicolumn{3}{r}{\textit{Continued on next page}} \\
\endfoot
 
\bottomrule
\endlastfoot
 
\multirow{5}{0.20\textwidth}{Build, Compilation, and Infrastructure Optimization}
 & Enable Build-Time Optimizations
 & Use explicit, unambiguous module paths so bundlers can resolve dependencies deterministically and apply stronger tree-shaking. \\
\cmidrule(l){2-3}
 & Performance-Optimized Dependency Selection
 & Upgrade libraries or frameworks to newer versions with faster compilers or more efficient runtime behavior. \\
\cmidrule(l){2-3}
 & Compiler Optimization Hints/Directives
 & Add explicit compiler hints (e.g., inlining) on verified hot paths to guide code generation beyond static analysis. \\
\cmidrule(l){2-3}
 & Reduce Rebuilding Size
 & Organize the build so only changed modules are recompiled, avoiding full rebuilds for small edits. \\
\cmidrule(l){2-3}
 & Enable Lazy Loading for Resources
 & Defer loading or initializing a resource until it is actually needed. \\
\midrule
 
\multirow{4}{0.20\textwidth}{Network, Database, and Data Access Optimization}
 & Batch API Requests (N+1)
 & Replace N sequential per-item queries with a single batched query, avoiding the N+1 problem. \\
\cmidrule(l){2-3}
 & Minimize Payload Size / Pass Reference Instead of Value
 & Send lightweight references (IDs) instead of full nested objects; fetch details only when needed. \\
\cmidrule(l){2-3}
 & Lazy Loading / Deferred Initialization
 & Defer fetching or constructing large datasets until first use, rather than loading everything upfront. \\
\cmidrule(l){2-3}
 & Relational Query Consolidation (N+1)
 & Replace multiple sequential relationship queries with a single query using JOINs or includes. \\
 
\end{longtable}
\endgroup

\begingroup
\setlength{\tabcolsep}{5pt}
\small
\begin{longtable}{L{0.68\textwidth} r}
\caption{Optimization-pattern catalog used for RQ2, with the number of patterns in each high-level category. The refined catalog contains 59 patterns across nine categories; Table~\ref{tab:new-patterns} details the patterns in the two newly introduced categories.}
\label{tab:pattern-catalog} \\
\toprule
\textbf{High-level category} & \textbf{Number of patterns} \\
\midrule
\endfirsthead
\multicolumn{2}{c}{\tablename\ \thetable{} -- continued from previous page} \\
\toprule
\textbf{High-level category} & \textbf{Number of patterns} \\
\midrule
\endhead
\bottomrule
\endlastfoot
Memory and Data Locality Optimizations & 12 \\
Code Smells and Structural Simplification & 8 \\
Control-Flow and Branching Optimizations & 7 \\
Algorithm-Level Optimizations & 6 \\
Loop Transformations & 6 \\
I/O and Synchronization & 7 \\
Build \& Compilation \& Infrastructure Optimization & 5 \\
Data Structure Selection and Adaptation & 4 \\
Network, Database, and Data Access Optimization & 4 \\
\midrule
\textbf{Total} & \textbf{59} \\
\end{longtable}
\endgroup

\subsubsection{Labeling Procedure}
\label{sec:methodology-rq2-labeling}
We use LLMs to assign optimization labels and use iterative human inspection to calibrate the annotation procedure~\citep{wang2024human}.

\myparagraph{Pilot study}
The pilot study used two commercial LLMs, \textsc{GPT-5.1} and \textsc{Gemini-3-Pro-Preview}, with temperature set to zero, to classify each patch independently.
Labels on which the models agreed were accepted directly, while disagreements were reviewed and adjudicated by two human annotators~\citep{peng2026agents, ahmed2025can}.

This procedure had two limitations for the large-scale extension.
First, manually adjudicating every model disagreement does not scale readily to the substantially larger dataset analyzed here.
Second, agreement between the two models did not guarantee a correct label: a manual audit of a 10\% random sample of model-agreed cases in the pilot study estimated an error rate of 10\%~\citep{peng2026agents}.
These limitations motivate the revised consensus procedure used in the large-scale extension.

\myparagraph{Large-scale extension}
We obtain independent labels from three models and retain a label only when at least two models agree on the category-pattern pair, without subsequent human adjudication. \textsc{GPT-5.6-sol}, \textsc{Gemini-3.1-Pro-Preview}, and \textsc{Qwen3.8-27B} each assign one category and one pattern to every PR in the sample of 2{,}260, with temperature set to zero.
This procedure requires corroboration by a second model and applies a uniform decision rule to ambiguous cases while remaining feasible at the scale of the dataset. Our annotation prompt is shown below; the complete prompt is available in the replication package~\citep{perfPatchStudy}.

\myprompt{RQ2 Classification Prompt}{
I have a performance optimization commit with the following information. Please analyze it with the following goals \dots
\textbf{Task Instructions:}
\begin{enumerate}
  \item \textbf{Code Function Explanation}: briefly explain what the code is doing---what problem it solves and how it works.
  \item \textbf{Optimization Comparison}: compare the original and optimized versions to identify algorithmic changes, performance improvements, redundant code removal, and other noteworthy structural or stylistic changes\dots
  \item \textbf{Optimization Pattern Classification}: based on the overall nature of the optimized code, assign exactly one optimization category and one most representative pattern from the given taxonomy.
\end{enumerate}

\textit{Output Requirements}: Respond in valid JSON with the fields \texttt{explanation}, \texttt{optimization\_comparison}, \texttt{high\_level\_pattern}, and \texttt{sub\_pattern}.

\textit{Input}: Optimization pattern taxonomy; code/commit context (original and optimized code).

}

The unit of classification is the pair \texttt{(high\_level\_pattern, sub\_pattern)}. 
We retain a label only when at least two models select the same pair.
All three models select the same pair for 1{,}238 of the 2{,}260 PRs: 644 agentic and 594 human-authored PRs. 
Exactly two models agree for a further 845 PRs: 405 agentic and 440 human-authored PRs. 
The majority agreement rate is therefore $(1{,}238+845)/2{,}260=92.2\%$. 
The remaining 177 PRs have no majority pair and are excluded from RQ2, leaving 2{,}083 PRs: 1{,}049 agentic and 1{,}034 human-authored PRs, corresponding to 92.8\% and 91.5\% of the respective groups. 
The similar retention rates indicate that this exclusion is not materially imbalanced by author type.

Disagreement is concentrated at the finer level of the taxonomy: in more than half of the 177 unresolved cases, the models agree on the category but select different patterns.\footnote{We refer to PRs for which no label received votes from at least two of the three models as \emph{unresolved}.}
We manually inspect 20 randomly selected disagreed cases to inspect the sources of disagreement.
Approximately half span multiple categories, making the required single-label assignment ambiguous.
A further quarter concern low-level or front-end optimizations that are not well represented in the catalog.
The remaining cases are not clearly performance-related or otherwise do not fit the available patterns.
Thus, unresolved cases tend to occur at the conceptual boundaries of the catalog.

\subsubsection{Comparison Procedure}
\label{sec:methodology-rq2-analysis}
We compare the optimization-category distributions of agentic and human-authored PRs. 
Because both author type and optimization category are
categorical, we use a chi-square test to determine whether the overall distribution of optimization categories differs between the two groups, and
report Cram\'er's $V$ to quantify the size of the difference.
We also compare the observed coverage of the 59 individual patterns across agentic and human-authored PRs.

We then use the consensus RQ2 labels to examine whether the structural differences identified in RQ1 are concentrated in particular kinds of optimization. Specifically, we combine the labels with the PR-level measures defined in Section~\ref{sec:methodology-rq1} and compare changes in NLOC and function count across eight categories.\footnote{We exclude Loop Transformations because it contains only four agentic PRs.}
The structural-change measures are skewed, so we use Mann--Whitney $U$ tests to compare agentic and human-authored PRs within each category and report Cliff's $\delta$ as the effect size.
These comparisons involve multiple statistical tests, so we apply the Benjamini--Hochberg correction to control the false discovery rate and reduce the chance that some significant results arise only because many comparisons are performed.
We report the resulting adjusted $q$-values.

\subsection{Validation Evidence (RQ3)}
\label{sec:methodology-rq3}
To answer RQ3, we determine whether each PR reports validation of its performance claim and, when validation is reported, classify the primary form of evidence. We first define the validation taxonomy (\cref{sec:methodology-rq3-taxonomy}), then describe the labeling procedure (\cref{sec:methodology-rq3-labeling}), and finally specify the aggregate and temporal comparisons (\cref{sec:methodology-rq3-analysis}).

\subsubsection{Validation Evidence Taxonomy}
\label{sec:methodology-rq3-taxonomy}
The large-scale extension retains the pilot study’s validation-evidence taxonomy unchanged, preserving a consistent measure across both stages. 
Because PR artifacts do not record the development process itself, we analyze the validation described in those artifacts, rather than directly observing how it was performed.\footnote{Direct observation would require fine-grained activity data, such as those available in an IDE-instrumented dataset (\eg the MSR 2018 Mining Challenge~\citep{proksch2018enriched}).}
We treat explicit claims of testing or evaluation as evidence that these activities occurred, while acknowledging that a reported claim may be incomplete or inaccurate, particularly when generated by a model.

Following the taxonomy developed in the pilot study, we classify reported validation evidence as \emph{Benchmark-Based Validation}, \emph{Profiling-Based Validation}, \emph{Static-Reasoning-Based Validation}, or \emph{Anecdotal or Informal Local Testing}. 
The first three categories are defined in Section~\ref{sec:background-validation}; the fourth captures validation claims without quantitative measurements or analytical support.
PRs containing no evidence are labeled as having no performance validation.

\subsubsection{Labeling Procedure}
\label{sec:methodology-rq3-labeling}

\myparagraph{Pilot study}
The pilot study applied the same two LLMs used for RQ2 to evidence assembled from each PR's title, description, issue and review comments, commit messages, CI workflow names, and code diff.
The two models classified validation presence and type independently, achieving agreement of $\kappa=0.74$ for validation presence and $\kappa=0.75$ for validation type.
For the remaining PRs with model disagreement, two authors independently reviewed and adjudicated labels, with inter-annotator agreement of $\kappa=0.62$ and $\kappa=0.58$, respectively.
A manual audit of a 10\% sample of model-agreed cases estimated error rates of 6.9\% for validation presence and 10.0\% for validation type.

\myparagraph{Large-scale extension}
The large-scale extension uses the same evidence sources and validation categories as the pilot study, thereby preserving the construct being measured while adapting the annotation procedure to the larger sample.
Complete evidence could not be retrieved for two PRs; the RQ3 input therefore contains 2{,}258 of the 2{,}260 sampled PRs, comprising 1{,}129 PRs from each group.
We obtain independent classifications from three models as in RQ2.
Each model determines whether validation is reported and, when it is, assigns one primary validation type.
Our annotation prompt is shown below; the complete prompt is available in the replication package~\citep{perfPatchStudy}.

\myprompt{RQ3 Classification Prompt}{
You are a classifier for performance validation evidence in GitHub Pull Requests (PRs). Your job is to decide whether there is explicit performance validation, and if so, to classify it into exactly one \texttt{validation\_type} category \dots

\textbf{Task Instructions:}

\begin{enumerate}

  \item \textbf{Internalize the categories}: Benchmark-Based Validation (unit tests/microbenchmarks with before-and-after metrics); Profiling-Based Validation (stack samples, hotspots, flamegraphs); Static-Reasoning-Based Validation (reasoning about complexity/allocations without runtime evidence); Anecdotal/Informal Local Testing (claims with no quantitative evidence)\dots

  \item \textbf{Decide validation\_present}: set to TRUE only if explicit evidence is shown; never infer validation from performance intent alone.

  \item \textbf{Assign validation\_type}: if \texttt{validation\_present} is FALSE, set \texttt{validation\_type} to ``none'' with empty \texttt{evidence\_sources}; otherwise choose exactly one non-``none'' category and list where it is evidenced (pipeline, description, comments, code diff).

\end{enumerate}

\textit{Output Requirements}: Return strict JSON with exactly the keys \texttt{validation\_present}, \texttt{evidence\_sources}, \texttt{validation\_type}, \texttt{validation\_description}, \texttt{pipeline\_signal}, \texttt{description\_signal}, \texttt{comment\_signal}.

\textit{Input}: PR description, code diff, pipeline/CI signals, and comments.
}

Consensus is determined in two stages.
Stage~1 determines validation presence by majority vote.
Because each of the three models cast a Boolean vote, every PR receives a majority decision.
This stage identifies validation evidence in 1{,}839 PRs (929 agentic and 910 human-authored) and no validation evidence in 419 PRs (200 agentic and 219 human-authored).
Stage~2 considers only the 1{,}839 PRs with validation evidence and assigns a primary type when at least two models select the same category.
All three models agree for 1{,}396 PRs, and exactly two agree for a further 423, producing an agreement rate of 98.9\%.
The remaining 20 PRs (11 agentic and 9 human-authored) have no majority type and are excluded from the validation-type analysis.
The resulting sample contains 1{,}819 PRs: 918 agentic and 901 human-authored.

Manual review indicates that disagreements about validation presence arise primarily from statements describing performance impact.
A PR may present an expected improvement as the motivation for a change rather than as evidence that the change was evaluated, but some models nevertheless interpret such statements as validation evidence.

\subsubsection{Aggregate and Temporal Comparisons}
\label{sec:methodology-rq3-analysis}
We compare the two author types in terms of both validation presence and the distribution of primary validation types. Both are categorical outcomes, so we use chi-square tests to determine whether they differ by author type and report Cram\'er's $V$ to quantify the size of the association.

Aggregate comparisons across the 65 sampled ISO weeks may obscure changes over time. 
We therefore also examine whether two outcomes change during the observation period: whether validation is reported and, among PRs with a resolved evidence type, whether benchmark-based validation is the primary evidence. 
The 1:1 balance within each ISO week permits the groups to be
compared over the same calendar periods.

Both temporal outcomes are binary, so we use logistic regression to estimate how they change over time and whether that change differs between agentic and human-authored PRs. For each temporal outcome $Y_i$, we fit the following
model:
\begin{equation}
\operatorname{logit}\Pr(Y_i=1) = \beta_0 + \beta_1 A_i + \beta_2 T_i + \beta_3 A_i T_i,
\label{eq:rq3-temporal}
\end{equation}
where $A_i$ indicates an agentic PR and $T_i$ denotes the PR's creation time in years, centered on the sample mean. 
The interaction term $\beta_3$ captures whether the temporal trend differs between agentic and human-authored PRs.
We test this difference using a Wald test and cluster the standard errors by repository to account for PRs from the same repository not being fully independent.
For the quarterly proportions shown descriptively, we report 95\% Wilson confidence intervals. 
Wilson intervals are used because the number of PRs varies across quarters and they provide more reliable intervals for proportions than the standard normal approximation.

\subsection{Reported Performance Metrics (RQ4)}
\label{sec:methodology-rq4}
RQ4, introduced in the large-scale extension, examines the quantitative content of the validation evidence identified in RQ3. For PRs with reported validation and an available RQ2 optimization label, we identify any quantitative performance metrics in the PR artifacts and compare them with the performance dimensions associated with the assigned optimization pattern. The analysis is limited to reported evidence; we do not execute the changed code, repeat benchmarks or profiling, or independently verify reported values or performance claims.
We extract reported metrics (Section~\ref{sec:methodology-rq4-metrics}), examine how they relate to the applied optimization and vary across optimization categories and evidence types (Section~\ref{sec:methodology-rq4-mapping}), and assess runtime--memory trade-off reporting and differences in validation practices between code-smell refactorings and other optimizations (Section~\ref{sec:methodology-rq4-tradeoffs-refactorings}).

\subsubsection{Metric Extraction}
\label{sec:methodology-rq4-metrics}
We use regular expressions to extract the quantitative performance metrics reported in each PR's validation evidence. Table~\ref{tab:performance-dimensions} defines nine metrics (M1--M9), and we additionally add a residual category (M0) for quantified gains that name no specific metric, such as ``3--5x overall performance improvement.''
Extraction is multi-label: a PR may report no metric, one metric, or several metrics.\footnote{We extract quantitative performance claims only. Static properties such as asymptotic complexity are represented in RQ3 as static-reasoning-based validation rather than as reported metrics.} For each metric, the extractor requires both a metric-specific keyword (\eg \texttt{throughput}, \texttt{heap}, or \texttt{bundle size}) and a quantitative expression, either a number with an appropriate unit (\eg \texttt{120ms}, \texttt{4MB}, or \texttt{1200 req/s}) or a comparative statement (\eg \texttt{2.5x faster}, \texttt{reduced from 125ns to 83ns}, or \texttt{-40\%}). M0 is assigned only when a quantified performance gain cannot be associated with a named metric.

\myparagraph{Extraction accuracy}
The extractor can miss evidence presented only in figures, screenshots, or linked dashboards, and its regular expressions can produce false positives. To assess precision, we use the same three LLMs employed in RQ3 to review a random 10\% sample of the 878 PRs for which the extractor identifies at least one metric, balanced between 44 agentic and 44 human-authored PRs. The audit identifies 77 true positives and 11 false positives (12.5\%). False positives commonly occur when a numeric configuration value, such as a memory limit, describes the evaluation environment rather than a performance result.

We perform two additional consistency checks. First, the extractor identifies a quantitative metric in 42 of the 382 PRs that RQ3 classifies as having no validation evidence (11.0\%; 11.6\% of agentic and 10.4\% of human-authored PRs). These disagreements flag possible false positives, although they could also reflect errors in the RQ3 labels. Second, we compare detection rates across RQ3 evidence types, expecting extractable metrics to be more common in benchmark-based evidence than in static reasoning (Section~\ref{sec:results-rq4}). We do not estimate recall because we do not audit PRs for metrics presented only in non-textual or externally linked evidence.

\subsubsection{Reported Metrics and Optimization Correspondence}
\label{sec:methodology-rq4-mapping}
We first characterize metric reporting separately by author type. For each PR, we count the number of distinct metrics reported; across PRs, we calculate the frequency of each metric. We then assess whether the reported metrics correspond to the expected benefits of the applied optimization. For each catalog pattern, we map the catalog's \emph{optimized metrics}~\citep{peng2026sysllmatic} to the taxonomy in Table~\ref{tab:performance-dimensions} and determine whether a PR reports at least one metric in the resulting set.
To determine whether the observed correspondence exceeds what would be expected by chance, we use a one-sided, 20{,}000-iteration permutation test.
The permutation procedure reassigns the expected-metric sets among optimization patterns while holding each PR's reported metrics fixed, thereby preserving the observed metric-reporting distribution while breaking the mapping between
optimization patterns and their expected metrics. 
We use a one-sided test because the hypothesis of interest is whether correspondence is greater than the permutation baseline.

Finally, we examine how metric reporting varies by optimization category and validation-evidence type, separately for agentic and human-authored PRs. For each RQ2 category, we construct a \emph{metric profile} comprising the frequency of every reported metric. We also compare the proportion of PRs reporting at least one metric in benchmark-based evidence with the corresponding proportion in each other RQ3 evidence type. 
Metric presence is binary, so these comparisons use contingency-table tests.
We use a chi-square test when expected cell counts are adequate and Fisher's exact test otherwise. To account for the multiple evidence-type comparisons, we apply the Benjamini--Hochberg correction to control the false discovery
rate and report adjusted $q$-values.
This comparison provides the evidence-type consistency check described in Section~\ref{sec:methodology-rq4-metrics}.

\subsubsection{Runtime--Memory Trade-off Reporting and Code-Smell Refactorings}
\label{sec:methodology-rq4-tradeoffs-refactorings}
\myparagraph{Runtime--Memory Trade-off Reporting}
We examine whether validation evidence for caching and buffering reports both a runtime benefit and the associated memory cost.\footnote{We restrict this analysis to caching and buffering because their exchange of additional memory for reduced execution time or I/O is well established, whereas the trade-offs associated with other catalog patterns depend more strongly on their implementation and workload.}
The analysis proceeds in two stages. First, among PRs assigned either pattern and reporting benchmark- or profiling-based evidence, we use the regular-expression extractor described in Section~\ref{sec:methodology-rq4-metrics} to identify PRs that report both a runtime-side metric---execution time, throughput, or I/O (M1, M2, or M5)---and memory (M3). We calculate the prevalence of runtime-metric reporting and runtime--memory co-reporting separately by author type. Co-reporting alone does not establish a trade-off, because the reported memory effect may be either adverse or beneficial.

Second, we submit each PR that reports both dimensions to three LLMs for independent classification. Each model determines whether the reported evidence describes a \emph{runtime--memory trade-off}, in which runtime improves while memory worsens, or a \emph{joint improvement}, in which both runtime and memory improve. We assign the final classification by majority vote and compare the distribution of these classifications between agentic and human-authored PRs.

\myparagraph{Code-Smell Refactorings}
We next examine whether validation practices differ between code-smell refactorings and changes that directly target a performance resource. The RQ2 category \emph{Code Smells and Structural Simplification} includes changes whose primary effect is structural, such as removing redundant computation, dead code, or duplicated work. The expected direction of these changes may be more readily inferred from the code than that of resource-targeting optimizations, whose effects frequently depend on the workload or involve trade-offs (Section~\ref{sec:background-tradeoff}). Authors may therefore apply different standards of evidence to the two kinds of change.

We compare code-smell refactorings with the eight remaining optimization categories on three binary outcomes: whether a PR reports validation evidence; whether a validated PR reports at least one quantitative metric; and whether a PR both reports validation and includes at least one metric.
We first compare code-smell refactorings with the other performance changes separately for agentic and human-authored PRs. 
Since each outcome is binary, we use chi-square tests to determine whether its prevalence differs between the two types of changes.
We then examine whether the difference between code-smell refactorings and other performance changes itself varies by author type. 
For each binary outcome $Y_i$, we fit the following logistic regression:
\begin{equation}
\operatorname{logit}\Pr(Y_i=1) = \gamma_0 + \gamma_1 C_i + \gamma_2 A_i + \gamma_3 C_i A_i,
\label{eq:rq4-cleanups}
\end{equation}
where $C_i$ indicates a code-smell refactoring and $A_i$ indicates an agentic PR.
The interaction term $\gamma_3$ captures whether the contrast between code-smell refactorings and other performance changes differs by author type.
We evaluate this interaction using a likelihood-ratio test that compares the model containing the interaction with a model that does not contain it.
To account for the multiple related comparisons, we apply the
Benjamini--Hochberg correction to control the false discovery rate and report adjusted $q$-values.

\section{Results}
We report the results for the four research questions in sequence: adoption outcomes, patch size, and structural indicators of maintainability (\Cref{sec:result-rq1}); optimization patterns (\Cref{sec:result-rq2}); reported validation evidence (\Cref{sec:result-rq3}); and the performance metrics reported by that evidence (\Cref{sec:results-rq4}).

\subsection{Adoption, Patch Size, and Maintainability (RQ1)}
\label{sec:result-rq1}
We compare adoption outcomes, patch size, and structural maintainability indicators between agentic and human-authored performance PRs. 

\begin{figure}[htbp]
    \centering
    \includegraphics[width=0.78\linewidth]{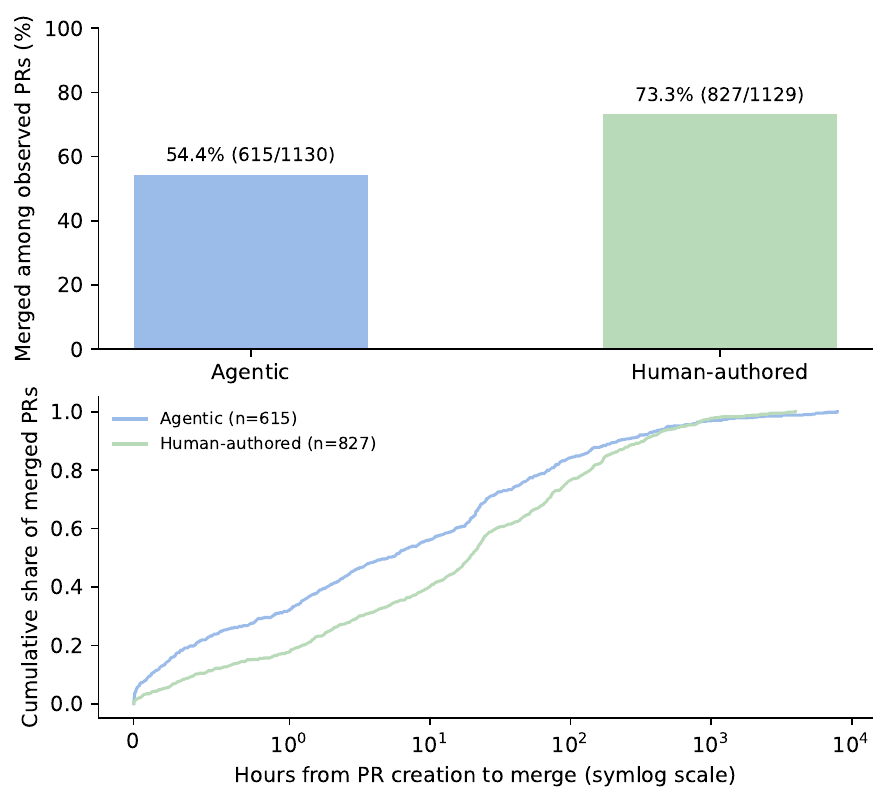}
    \caption{
    Adoption outcomes for agentic and human-authored performance PRs.
    Top: share of sampled PRs merged, with merged and observed counts.
    Bottom: cumulative distribution of time from PR creation to merge among merged PRs, on a symlog hour axis.
    Agentic PRs are merged less often, but a larger share of merged agentic PRs is merged within the first hours (median 4.99 vs.\ 19.16 hours).
    }
    \label{fig:merge-outcomes}
\end{figure}

\myparagraph{Adoption Outcomes}
As shown in Figure~\ref{fig:merge-outcomes}, agentic performance PRs have a lower merge rate than human-authored PRs (54.4\% vs.\ 73.3\%; $q<0.001$, $V=0.195$).
Among merged PRs, however, agentic PRs have a shorter median time to merge (4.99 vs.\ 19.16 hours; $q<0.001$, $\delta=-0.207$).
The cumulative distributions show that a larger share of merged agentic PRs reach merge within the first few hours; the curves draw closer among PRs that take longer to merge.\footnote{This timing result applies only to merged PRs and does not establish that agentic PRs receive faster review overall.}

\begin{figure}[htbp]
    \centering
    \includegraphics[width=\linewidth]{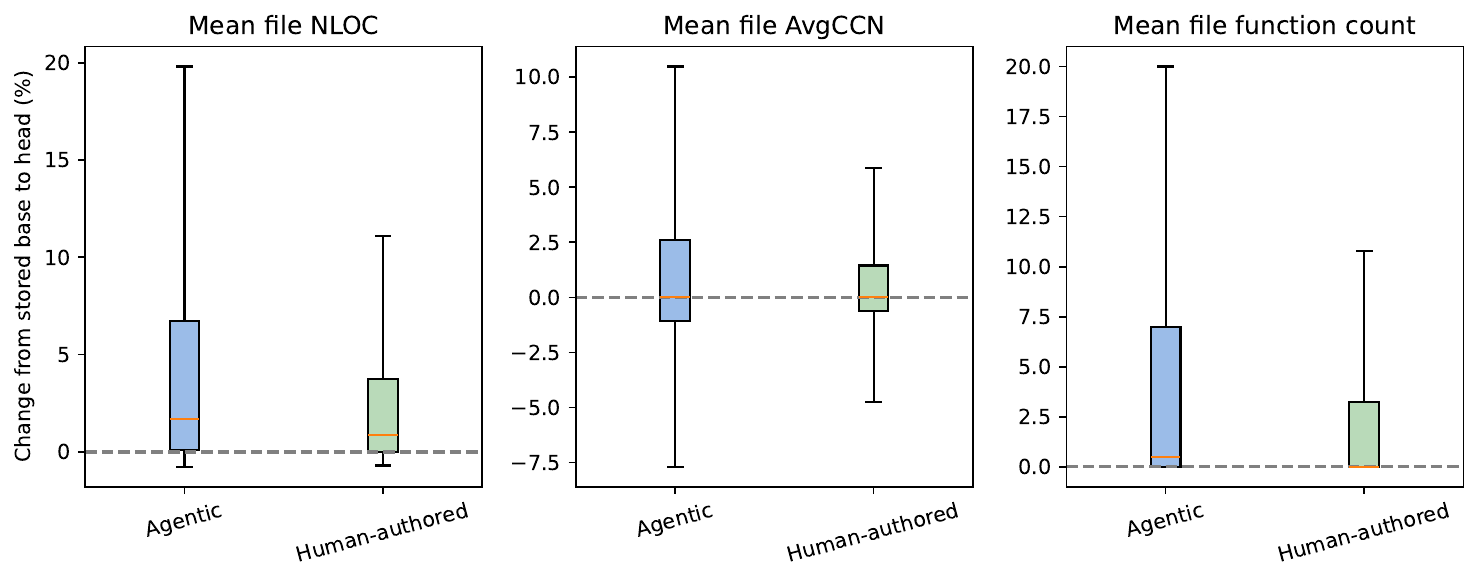}
    \caption{
    Per-PR percentage change from base to head in three structural indicators of maintainability, computed with Lizard~\citep{yin2019lizard} over each PR's changed files: mean file NLOC, mean file average cyclomatic complexity (AvgCCN), and mean file function count ($n=2{,}029$ PRs).
    Boxes show medians and interquartile ranges, whiskers the 10th and 90th percentiles, and the dashed line no change.
    The distributions overlap substantially and the medians lie near zero, but agentic PRs show wider spread and heavier positive tails in NLOC and function count, while AvgCCN does not differ between groups.
    }
    \label{fig:maintainability}
\end{figure}

\myparagraph{Patch Size and Maintainability Indicators}
Agentic PRs have more changed lines and commits than human-authored PRs: the median number of added plus deleted lines is 130.5 versus 81.0, and the median commit count is three versus two ($q<0.001$; $\delta=0.113$ and $0.097$, respectively). Both groups change a median of three files ($q=0.222$). Among the 2{,}029 cases with structural measures, agentic PRs also show slightly larger median increases in NLOC (1.70\% vs.\ 0.87\%) and function count (0.50\% vs.\ 0.00\%; $q<0.001$ for both; $\delta=0.120$ and $0.110$, respectively), while average cyclomatic complexity does not differ ($q=0.624$, $\delta=0.013$). As Figure~\ref{fig:maintainability} shows, the distributions of all three indicators overlap substantially. Agentic PRs nevertheless exhibit heavier positive tails, suggesting that a subset of these PRs produces comparatively large structural increases, which we further investigate in Section~\ref{sec:result-rq2}.

\rqconclusion{RQ1: How do agentic and human-authored performance PRs differ in adoption outcomes, patch size, and structural indicators of maintainability?}{Agentic performance PRs are merged less often than human-authored PRs (54.4\% vs.\ 73.3\%) but, among merged PRs, have a shorter median time to merge (4.99 vs.\ 19.16 hours). Their patches contain more changed lines (median 130.5 vs.\ 81.0) and commits (3 vs.\ 2). Structural differences are small: median NLOC growth is 1.70\% versus 0.87\% ($\delta=0.120$), median function-count growth is 0.50\% versus 0.00\% ($\delta=0.110$), and average cyclomatic-complexity changes do not differ ($q=0.624$).}

\subsection{Optimization Patterns (RQ2)}
\label{sec:result-rq2}
To answer RQ2, we first compare the optimization categories and individual patterns represented in agentic and human-authored performance PRs.
We then use these labels to examine whether the aggregate structural difference identified in RQ1 is concentrated in particular optimization categories.

\subsubsection{Optimization Categories and Pattern Coverage}
\begin{figure}[htbp]
    \centering
    \includegraphics[width=\linewidth]{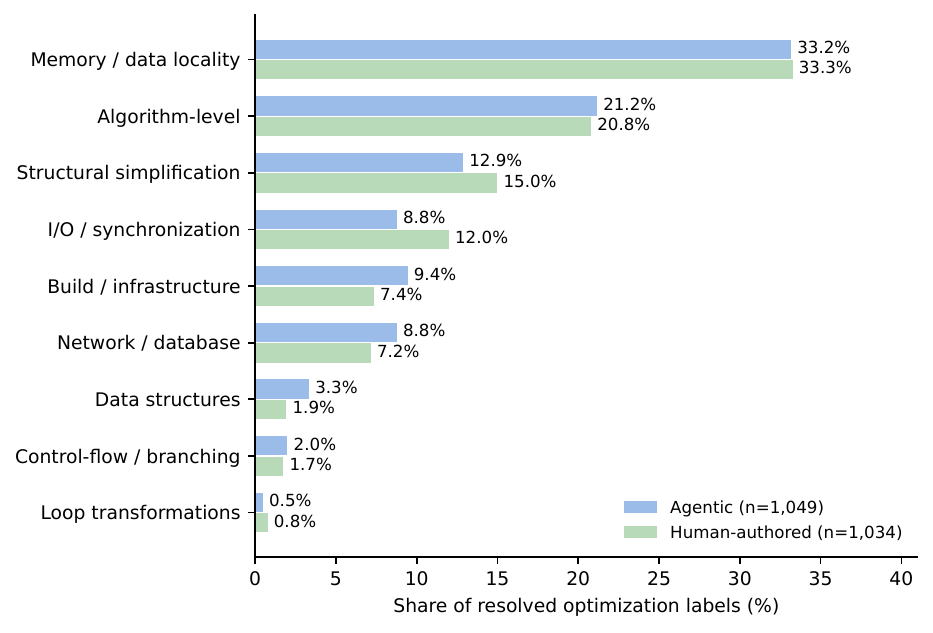}
    \caption{
    Distribution of the nine optimization categories among the 2{,}083 PRs with a majority consensus label (1{,}049 agentic and 1{,}034 human-authored).
    Percentages are computed within each author group and sum to 100\% per group.
    The two distributions are nearly identical: memory and data-locality optimizations and algorithm-level optimizations account for more than half of each group, and no category differs by more than 3.2 percentage points.
    }
\label{fig:pattern_distribution}
\end{figure}

Figure~\ref{fig:pattern_distribution} shows similar category distributions for the two groups.
Memory and data-locality optimizations are the most prevalent category (33.2\% of agentic PRs and 33.3\% of human-authored PRs), followed by algorithm-level optimizations (21.2\% and 20.8\%, respectively).
Author type is associated with optimization category ($\chi^2(8)=16.14$, $p=0.040$), but the association is small (Cram\'er's $V=0.088$).
The largest gap occurs in I/O and synchronization, which accounts for 8.8\% of agentic PRs and 12.0\% of human-authored PRs, a difference of 3.2 percentage points.
At the finer-grained pattern level, agentic PRs represent 44 of the 59 catalog patterns, while human-authored PRs represent 43.
Thus, \emph{the two groups have nearly identical observed pattern coverage in the balanced sample}.

\myparagraph{Structural Change within Optimization Categories}
We next examine whether the greater aggregate NLOC growth of agentic PRs reported in Section~\ref{sec:result-rq1} occurs across optimization categories or is localized to particular types of work.
Consensus optimization labels and per-PR NLOC measures are available for 1{,}905 of the 2{,}083 labeled PRs (91.5\%); the remaining PRs modify no files that the metric tooling can parse.
The between-group differences are concentrated in two categories.
For build and infrastructure work, the median percentage change in mean file NLOC is 2.43\% for agentic PRs and 0.21\% for human-authored PRs ($q=0.0004$, $\delta=0.403$); the corresponding median absolute changes are 8.60 and 1.00 lines ($q=0.0006$, $\delta=0.392$).
For network, database, and data-access work, the median percentage changes are 2.74\% and 1.05\% ($q=0.041$, $\delta=0.241$), and the median absolute changes are 14.83 and 5.45 lines ($q=0.013$, $\delta=0.276$).
No other category exhibits a statistically significant difference.
This result shows that \emph{the greater NLOC growth of agentic PRs is not distributed across optimization patterns generally but is concentrated in infrastructure and data-access optimizations}.

\rqconclusion{RQ2: Which optimization patterns occur in agentic and human-authored performance PRs, and how does their prevalence differ between the two groups?}{Agentic and human-authored PRs draw on a similar repertoire of optimization strategies: the category-level association is small ($V=0.088$), and the groups cover 44 and 43 catalog patterns, respectively. The greater NLOC growth of agentic PRs is localized to build and infrastructure optimizations (median 2.43\% vs.\ 0.21\%, $\delta=0.403$) and to network, database, and data-access optimizations (median 2.74\% vs.\ 1.05\%, $\delta=0.241$).}

\subsection{Validation Evidence (RQ3)}
\label{sec:result-rq3}
To answer RQ3, we compare agentic and human-authored performance PRs in terms of whether they report validation evidence and which form of evidence they report as primary.
We first present the aggregate comparison, then examine temporal changes in both outcomes.

\subsubsection{Validation Presence and Type}
\myparagraph{Validation Presence}
Validation evidence is reported in 929 of 1{,}129 agentic PRs (82.3\%) and 910 of 1{,}129 human-authored PRs (80.6\%).
Author type is not significantly associated with validation presence ($\chi^2(1)=0.95$, $p=0.330$; Cram\'er's $V=0.021$).
The aggregate comparison therefore provides \emph{no evidence that agentic and human-authored PRs differ in whether validation is reported}.
However, the aggregate rates mask temporal variation: during the first half of the sampling window, agentic PRs report validation less often than human-authored PRs.
This earlier gap is consistent with the pilot study, whose sample falls within this earlier period and which reported validation less often among agentic PRs (45.7\% vs.\ 63.6\%)~\citep{peng2026agents}.
Section~\ref{sec:result-rq3-temporal} examines how the relationship between author type and reported validation changes over time.

\begin{figure}[htbp]
    \centering
    \includegraphics[width=1\linewidth]{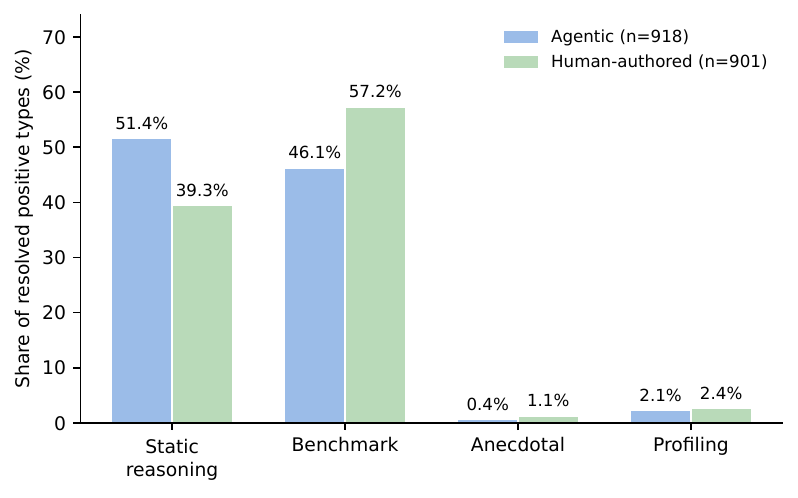}
    \caption{
    Primary validation-evidence type among the 1{,}819 PRs with reported validation evidence and a resolved evidence type (918 agentic and 901 human-authored).
    Percentages are computed within each author group and sum to 100\% per group.
    Static reasoning and benchmarks together account for more than 96\% of evidence in both groups, but in opposite proportions: static reasoning leads among agentic PRs, whereas benchmark-based evidence leads among human-authored PRs. Profiling and anecdotal evidence are rare in both.
    }
    \label{fig:validation_type_by_author}
\end{figure}

\myparagraph{Validation Type}
Among the 1{,}819 PRs with reported validation evidence and evidence type, the distribution of evidence types differs by author ($\chi^2(3)=28.52$, $p<0.001$; Cram\'er's $V=0.125$).
As shown in Figure~\ref{fig:validation_type_by_author}, static reasoning is the evidence more often in agentic PRs than in human-authored PRs (51.4\% vs.\ 39.3\%), whereas benchmark-based evidence is more frequent in human-authored PRs (57.2\% vs.\ 46.1\%).
Profiling is in 2.1\% of agentic PRs and 2.4\% of human-authored PRs, while anecdotal evidence is in 0.4\% and 1.1\%, respectively. Together, these two forms account for fewer than 4\% of these PRs in either group.
Thus, \emph{agentic and human-authored PRs report validation at similar rates but differ in its form: agentic PRs rely more heavily on static reasoning, whereas human-authored PRs rely more heavily on benchmarks}.

\subsubsection{Validation over Time}
\label{sec:result-rq3-temporal}
\begin{figure}[htbp]
    \centering
    \includegraphics[width=1\linewidth]{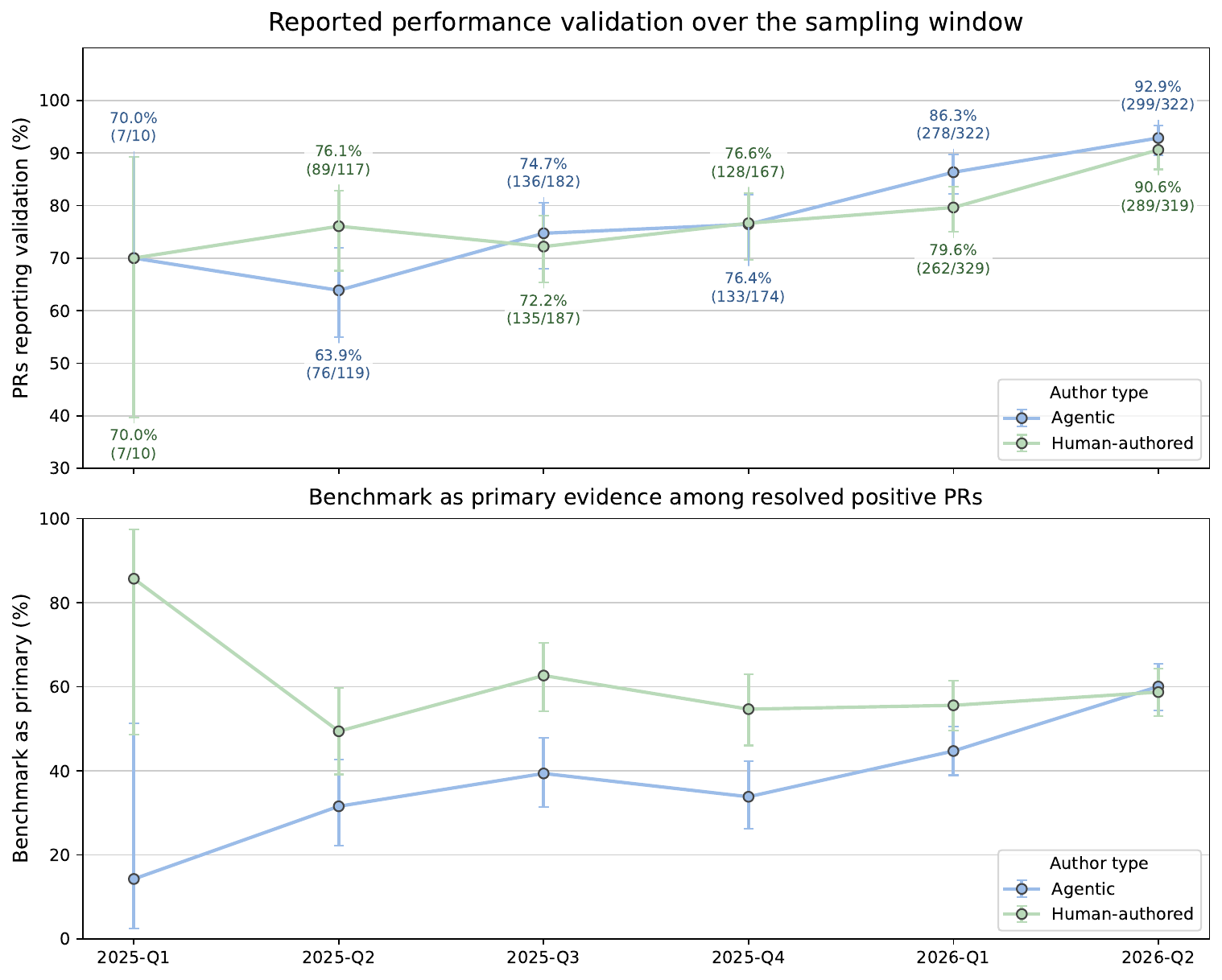}
    \caption{
    Quarterly trends in reported validation, with $95\%$ Wilson intervals. The top panel shows the share of performance PRs reporting any validation evidence. The bottom panel shows, among PRs with reported validation evidence and evidence type, the share whose primary evidence is benchmark-based.
    Validation prevalence increases in both groups, whereas the benchmark share increases among agentic PRs and remains stable among human-authored PRs.
    }
    \label{fig:validation-temporal}
\end{figure}

\myparagraph{Validation Presence over Time}
Figure~\ref{fig:validation-temporal} shows that reported validation increases from approximately 70\% to more than 90\% over the study period in both groups.
In the validation-presence model, the odds of reporting validation increase by a factor of $3.06$ per year for human-authored PRs ($95\%$ CI $[1.71, 5.49]$) and $6.47$ per year for agentic PRs ($95\%$ CI $[3.30, 12.70]$; both $p<0.001$).
The observed between-group difference changes direction over time: agentic PRs report validation less often than human-authored PRs in the first half of the study period (69.5\% vs.\ 74.1\%), but more often in the second half (88.0\% vs.\ 83.5\%).
The repository-clustered group-by-time interaction is not statistically significant ($p=0.093$). We therefore interpret the crossover descriptively rather than as evidence that the groups follow different temporal trends. The increase observed in both groups is consistent with a period effect shared across the sampled repositories.

\myparagraph{Validation Type over Time}
The form of reported evidence follows a different temporal pattern.
Among the 1{,}819 PRs with reported validation evidence and evidence type, the share of human-authored PRs using benchmark-based evidence remains stable (odds ratio per year $=0.98$, $95\%$ CI $[0.56, 1.73]$, $p=0.950$), whereas the corresponding share of agentic PRs increases (odds ratio per year $=3.63$, $95\%$ CI $[1.32, 9.96]$, $p=0.012$).
The difference between these temporal trends remains statistically significant after clustering standard errors by repository (interaction odds ratio $=3.69$, $95\%$ CI $[1.24, 10.97]$, $p=0.019$).
At the quarterly level, benchmark-based evidence is primary in 31.6\% of agentic PRs with reported validation evidence and evidence type in 2025-Q2, compared with 49.4\% of the corresponding human-authored PRs. By 2026-Q2, the respective proportions are 60.1\% and 58.7\%.
Thus, \emph{reported validation becomes more prevalent in both groups. The use of benchmark-based evidence follows a different temporal pattern by author type, increasing among agentic PRs while remaining stable among human-authored PRs}.

The evidence-type labels describe the reported form of validation.
They do not establish whether the evidence is correct, reproducible, or sufficient to support the associated performance claim.
A benchmark-based label therefore establishes only that a measurement is reported, not that the measurement was produced by a sound experiment or can be reproduced.
RQ4 therefore examines whether the reported evidence includes any quantitative performance metrics and, if so, which metrics are reported (Section~\ref{sec:results-rq4}).

\rqconclusion{RQ3: How do agentic and human-authored performance PRs differ in their reported testing and validation practices, and how do these practices change over time?}{Agentic and human-authored PRs report validation at similar overall rates (82.3\% vs.\ 80.6\%, $p=0.330$) but differ in its form: static reasoning is more frequent among agentic PRs (51.4\% vs.\ 39.3\%), whereas benchmark-based evidence is more frequent among human-authored PRs (57.2\% vs.\ 46.1\%). Benchmark-based evidence increases over time among agentic PRs (odds ratio per year $=3.63$, $p=0.012$) but remains stable among human-authored PRs (odds ratio $=0.98$, $p=0.950$); by 2026-Q2, the corresponding shares are 60.1\% and 58.7\%.}

\subsection{Reported Performance Metrics (RQ4)}
\label{sec:results-rq4}
RQ4 characterizes the quantitative performance claims contained in the validation evidence reported in PR artifacts.\footnote{We do not execute the changed code, reproduce the reported experiments, or independently verify the performance claims.}
The analysis covers the 1{,}699 PRs that RQ3 classifies as reporting validation evidence and for which an RQ2 optimization label is available (858 agentic and 841 human-authored).
We first examine the number and types of metrics reported, their correspondence with the applied optimization, and their distribution across optimization categories and evidence types (Section~\ref{sec:results-rq4-metrics}).
We then examine whether the reported evidence captures runtime--memory trade-offs in caching and buffering changes and whether code-smell refactorings report validation evidence and quantitative metrics at the same rates as other performance changes (Section~\ref{sec:results-rq4-focused}).

\subsubsection{Metric Reporting and Optimization Correspondence}
\label{sec:results-rq4-metrics}
\begin{figure}[htbp]
    \centering
    \includegraphics[width=0.7\linewidth]{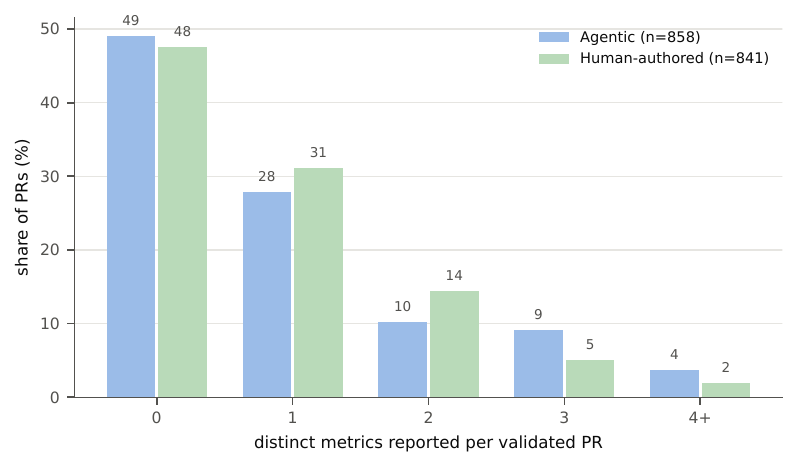}
    \caption{
    Number of distinct quantitative metrics extracted per PR with validation evidence (858 agentic and 841 human-authored PRs).
    Bars show the share of each group's PRs with validation evidence reporting zero, one, two, three, or four or more metrics.
    About half of PRs with validation evidence in both groups report no metric, and among those that do, a single metric is the most common case; agentic PRs are somewhat more likely to report three or more.
    }
    \label{fig:rq4-dimensionality}
\end{figure}

\myparagraph{Number of Metrics}
We begin by counting the distinct quantitative metrics reported in each PR with validation evidence.
At least one metric is identified in 437 of the 858 agentic PRs (50.9\%) and 441 of the 841 human-authored PRs (52.4\%).
Among the 878 PRs reporting at least one metric, 501 (57.1\%) report exactly one: 239 agentic and 262 human-authored PRs.
Two or more metrics are reported in 198 agentic PRs (23.1\% of agentic PRs with validation evidence) and 179 human-authored PRs (21.3\% of human-authored PRs with validation evidence) (Fig.~\ref{fig:rq4-dimensionality}).
These results indicate that \emph{half of the PRs with reported validation contain no extractable quantitative performance metric, and a majority of those that report a metric report only one}.

\begin{figure}[htbp]
    \centering
    \includegraphics[width=\linewidth]{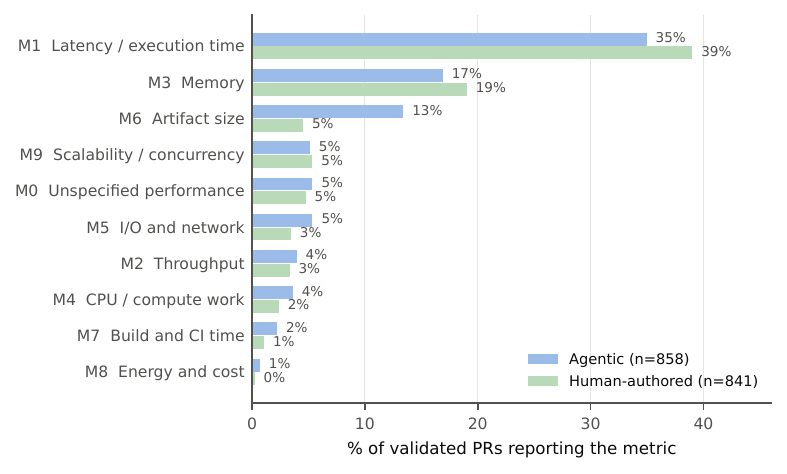}
    \caption{
    Share of PRs with validation evidence that report each performance metric from Table~\ref{tab:performance-dimensions}, by author type (858 agentic and 841 human-authored PRs).
    Extraction is multi-label, so a PR can contribute to several bars and shares do not sum to 100\%.
    Latency and execution time (M1) lead in both groups, followed by memory (M3); the largest between-group gap is artifact size (M6), reported almost three times as often in agentic PRs. Every other metric is reported by 5\% or fewer of PRs in either group.
    }
    \label{fig:rq4-metric-frequency}
\end{figure}

\myparagraph{Metrics Reported}
Figure~\ref{fig:rq4-metric-frequency} shows that latency and execution time are the most frequently reported metrics in both groups, appearing in 300 agentic PRs (35.0\%) and 328 human-authored PRs (39.0\%).
Memory is reported in 145 agentic PRs (16.9\%) and 160 human-authored PRs (19.0\%), while artifact size is reported in 115 (13.4\%) and 38 (4.5\%), respectively.
No other metric reaches 6\% in either group.
Scalability, I/O and network activity, throughput, CPU use, and build time are each reported in 1--5\% of PRs with validation evidence, while approximately 5\% report a quantified gain without identifying a specific metric (M0).
Energy or cost is reported in only six agentic and two human-authored PRs, representing less than 1\% of each group.

\myparagraph{Optimization Correspondence}
We next assess whether the reported metrics correspond to the expected effects of the applied optimization pattern, as specified in the catalog.
Among the 1{,}699 PRs with validation evidence, 34.5\% report at least one metric that the assigned pattern is expected to improve, compared with 23.7\% under the permutation baseline ($p=0.002$).
The observed proportion also exceeds the permutation baseline within both author groups (32.3\% vs.\ 23.5\% for agentic PRs; 36.7\% vs.\ 24.6\% for human-authored PRs) and within each evidence type.
Thus, \emph{reported metrics are associated with the expected effects of the applied optimization rather than being independent of the optimization pattern}.

\begin{figure}[htbp]
    \centering
    \includegraphics[width=\linewidth]{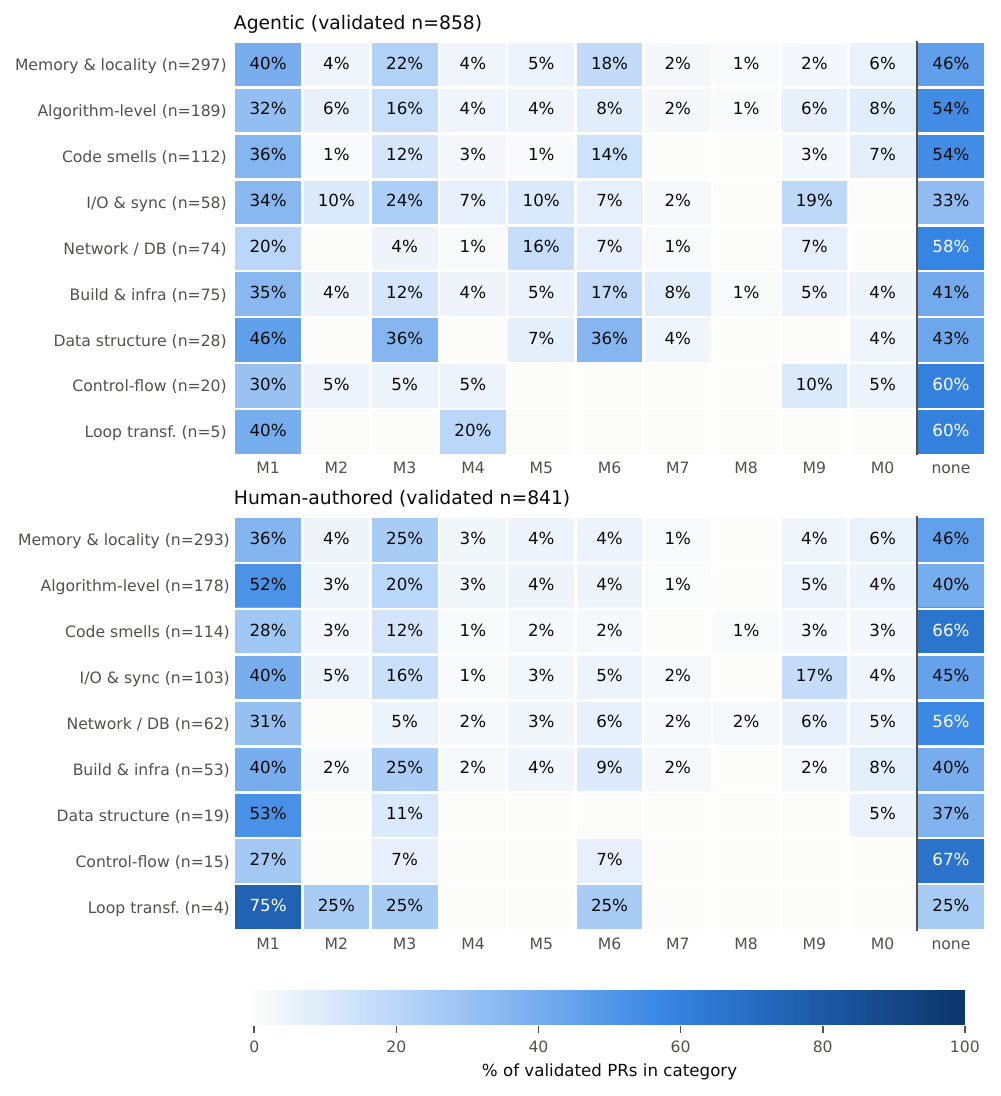}
    \caption{
    Metric profile of each optimization category, for agentic (top) and human-authored (bottom) PRs with validation evidence.
    Each cell gives the percentage of the category's PRs with validation evidence that report the metric in that column (as defined in Table~\ref{tab:performance-dimensions}); \textit{none} is the share with no extracted metric, and row counts give the number of PRs with validation evidence per category. Because extraction is multi-label, rows do not sum to 100\%.
    Latency (M1) is the leading metric in every category and both groups; the \textit{none} column is substantial throughout, and category-specific signals are limited to elevated scalability reporting (M9) for I/O and synchronization changes in both groups and elevated I/O and network reporting (M5) for network and data-access changes among agentic PRs.
    }
    \label{fig:rq4-metric-profile}
\end{figure}

\myparagraph{Metric Profile by Category}
Figure~\ref{fig:rq4-metric-profile} presents the metric profile of each optimization category.
Latency is the most frequently reported metric in every category and in both groups.
Beyond latency, metric reporting is sparse: most other metrics appear in substantially fewer than one-quarter of the PRs in a category, and several are absent from multiple categories.
Among agentic PRs, network, database, and data-access changes exhibit the clearest category-specific profile, with I/O and network metrics reported more frequently than in any other category; human-authored PRs in this category show no such concentration.
Although the proportion of PRs with validation evidence but no identified metric varies across categories, it remains substantial in every category.

\myparagraph{Metric Reporting by Validation Type}
As a consistency check, we compare metric-detection rates across the RQ3 evidence types, for which benchmark-based evidence is expected to contain quantitative metrics more often than the other forms of evidence (Section~\ref{sec:methodology-rq4-metrics}).
Among the 1{,}684 PRs with an evidence type, the pipeline identifies at least one metric in 79.8\% of benchmark-based PRs, 30.8\% of profiling-based PRs, 21.6\% of static-reasoning-based PRs, and 18.2\% of anecdotal ones.
The detection rate for benchmark-based evidence differs significantly from that for each other evidence type ($q<0.001$).
This distribution is \emph{consistent with the expected relationship between evidence type and quantitative metric reporting}.

\subsubsection{Runtime--Memory Trade-off Reporting and Code-Smell Refactorings}
\label{sec:results-rq4-focused}
\myparagraph{Runtime--Memory Trade-off Reporting}
We examine whether validation evidence for caching and buffering reports both runtime and memory effects, restricting the analysis to PRs with benchmark- or profiling-based evidence.
Among the resulting 130 PRs (74 agentic and 56 human-authored), 96 (73.8\%) report an improvement in execution time, throughput, or I/O; 29 of these 96 PRs also report a memory metric.
The proportion reporting a runtime-side improvement is similar for agentic and human-authored PRs (74.3\% vs.\ 73.2\%), as is the proportion reporting both runtime and memory metrics (23.0\% vs.\ 21.4\%).

LLM classification identifies 17 of the 29 co-reporting PRs as runtime--memory trade-offs; the remaining 12 report improvements in both dimensions.
The proportion classified as a trade-off is nearly identical across author types: 10 of 17 agentic PRs (58.8\%) and 7 of 12 human-authored PRs (58.3\%).
Relative to the 96 PRs reporting a runtime-side improvement, 17.7\% also report an adverse memory effect (18.2\% of agentic PRs and 17.1\% of human-authored PRs).
Thus, \emph{fewer than one in five caching or buffering PRs that report a runtime-side improvement also report an adverse memory effect, with nearly identical proportions for the two author types}.

\myparagraph{Code-Smell Refactorings vs.\ Performance Changes}
The \emph{Code Smells and Structural Simplification} category has the lowest overall rate of metric reporting in Fig.~\ref{fig:rq4-metric-profile}.
To determine whether this difference varies by author type, Table~\ref{tab:rq4-cleanups} compares these code-smell refactorings (290 PRs) with the eight categories that directly target a performance resource (1{,}791 PRs).
Among human-authored PRs, code-smell refactorings report validation less often than other performance changes (73.5\% vs.\ 82.8\%); among PRs with validation evidence, code-smell refactorings also report at least one metric less often (34.2\% vs.\ 55.3\%).
Consequently, 25.2\% of human-authored code-smell refactorings report both validation evidence and at least one metric, compared with 45.8\% of other human-authored performance changes; all three contrasts are statistically significant ($q\leq0.013$).
Among agentic PRs, code-smell refactorings and other performance changes have similar rates of reported validation evidence (83.0\% vs.\ 81.7\%) and metric-reporting rates among PRs with validation evidence (45.5\% vs.\ 51.7\%).
The corresponding proportions reporting both validation evidence and at least one metric are 37.8\% and 42.3\%, and none of the three contrasts is statistically significant ($q\geq0.33$).
The category-by-author interaction remains significant after correction for all three outcomes (last row of Table~\ref{tab:rq4-cleanups}).
When code-smell refactorings do report a metric, their metric distribution resembles that of other PRs, with latency reported most frequently.
Overall, \emph{the reporting of validation evidence and quantitative metrics varies by the nature of the change among human-authored PRs, whereas the corresponding differences among agentic PRs are not statistically significant}.

\begin{table}[htbp]
\centering
\small
\setlength{\tabcolsep}{3pt}
\caption{Validation and metric reporting for code-smell refactorings (\emph{Code Smells and Structural Simplification}) and for performance changes (the eight remaining categories), by author type.
\emph{Validation evidence} and \emph{validation evidence and $\geq$1 metric} are shares of all PRs in the row; \emph{$\geq$1 metric} is the share of PRs with validation evidence.
The final row tests whether the refactoring-versus-performance contrast differs between author types.
Human-authored PRs report validation evidence and metrics for refactorings less often than for performance changes, whereas agentic PRs treat the two alike.}
\label{tab:rq4-cleanups}
\begin{tabular}{@{}llccc@{}}
\toprule
\textbf{Author} & \textbf{Category} & \makecell{\textbf{Validation}\\ \textbf{evidence}} & \textbf{$\geq$1 metric} & \makecell{\textbf{Validation evidence}\\ \textbf{and $\geq$1 metric}} \\
\midrule
Agentic & Refactorings & 83.0\% & 45.5\% & 37.8\% \\
Agentic & Perf.\ changes & 81.7\% & 51.7\% & 42.3\% \\
human-authored & Refactorings & 73.5\% & 34.2\% & 25.2\% \\
human-authored & Perf.\ changes & 82.8\% & 55.3\% & 45.8\% \\
\midrule
\multicolumn{2}{@{}l}{Category $\times$ author interaction, $p$, $q$} & 0.043, 0.043 & 0.034, 0.043 & 0.007, 0.021 \\
\bottomrule
\end{tabular}
\end{table}

\rqconclusion{RQ4: What quantitative performance metrics, if any, are reported in the validation evidence of agentic and human-authored performance PRs, and how do those reported metrics relate to the applied optimization?}{Among PRs with validation evidence, only 50.9\% of agentic PRs and 52.4\% of human-authored PRs report an extractable performance metric; among the 878 PRs that do, 501 (57.1\%) report exactly one. Metric--optimization correspondence exceeds the permutation baseline (34.5\% vs.\ 23.7\%, $p=0.002$), but only 17 of 96 caching or buffering PRs reporting a runtime-side gain also report an adverse memory effect (17.7\%). Among human-authored PRs, 25.2\% of code-smell refactorings report both validation evidence and at least one metric, compared with 45.8\% of other performance changes; the corresponding proportions for agentic PRs are 37.8\% and 42.3\%.}

\section{Threats to Validity}
\label{sec:Threats}

We discuss construct, internal, conclusion, and external threats to validity~\citep{wohlin2012experimentation}.

\myparagraph{Construct Validity}
Merge outcomes reflect repository integration workflows and do not directly measure review effort or change quality.
NLOC, average cyclomatic complexity, and function count capture code size and control-flow structure, but not readability, duplication, or architectural impact.

Dataset inclusion partly relies on an automated classifier evaluated against 100 PRs labeled by one author; classification and reference-label errors may therefore affect the sample.
The accepted RQ2 and RQ3 consensus labels were not independently re-audited and may contain classification errors.
RQ3 also measures documented rather than directly observed validation and may omit unreported activities.
Conversely, a reported benchmark does not establish that it was executed as described or that its results were reported accurately.
Agentic systems may also learn to reproduce the form of human validation reports, including plausible benchmark claims, without a corresponding improvement in measurement practice or evidential reliability; our artifact-based analysis cannot distinguish such imitation from genuine execution.

RQ4 neither verifies performance effects nor reproduces experiments.
The metric extractor's audit estimates 87.5\% precision but not recall, so reported frequencies may include false positives and omit evidence in figures, screenshots, or linked artifacts.

\myparagraph{Internal Validity}
The observational design identifies associations rather than causal effects of agentic authorship.
Weekly stratification balances submission time but not repository, programming language, contributor population, or integration workflow, any of which may contribute to the observed associations.
Exclusions for unavailable code metrics or unresolved labels could also introduce bias if omitted PRs differ systematically from retained PRs, although retention rates are similar across author types.

\myparagraph{Conclusion Validity}
Repository dependence is addressed only by the clustered standard errors in the RQ3 temporal analysis; the remaining inferences are unadjusted.
The RQ4 correspondence result depends on the randomization assumptions of its permutation procedure.
We thus interpret statistical significance along with effect sizes and the assumptions underlying each analysis.

\myparagraph{External Validity}
Our findings are limited to performance PRs in AIDev's public GitHub repository frame, which applies a 100-star threshold, and to its five covered agents through June 1, 2026.
They may not generalize to less prominent or proprietary repositories, other agents, later agent versions, or settings without visible agent involvement.
Because the agentic group contains autonomously authored and submitted PRs, it does not represent human-controlled AI-assisted development.

\section{Discussion}
\subsection{Agents as a Path to Stronger Performance Validation}
Performance optimization has historically been a difficult software engineering task, in part because establishing whether a change actually improves performance requires careful measurement, appropriate workloads, and control of confounding factors. 
Our results show that these challenges persist in human-authored PRs: although validation is common, quantitative measurements and trade-off analysis remain sparse across both groups (Sections~\ref{sec:result-rq3} and~\ref{sec:results-rq4}). 
This finding suggests that matching current human practice should not be the ultimate goal for coding agents.
As agents take on a larger role in performance-critical development, they could raise the standard of practice by making rigorous measurement, trade-off analysis, and reproducible validation routine parts of the optimization process. 
Agentic systems should therefore be evaluated not only on whether they match current developer practices, but also on whether they strengthen them.

\subsection{Towards Next-Gen Code Optimization}
Our analysis shows that agentic PRs draw from a similar optimization-pattern space as human-authored PRs (Section~\ref{sec:result-rq2}), highlighting both the effectiveness of current LLMs in acquiring established performance knowledge and the limits of their practical behavior.
This observed similarity may reflect that agents are genuinely capable of applying the same complex optimizations as humans, or it may indicate that agents have learned from human-written artifacts what opportunities to recognize and thus reproduce human blind spots.
Our data do not directly measure the available optimization opportunities,
the result therefore demonstrates convergence in observed strategy, not equivalence in underlying optimization capability.
Distinguishing these possibilities will require comparisons on matched bottlenecks that vary the available optimization opportunities.

Looking within this shared strategy space reveals a further opportunity: loop transformations remain the least applied category (Fig.~\ref{fig:pattern_distribution}), despite decades of work showing the effectiveness of techniques such as unrolling~\citep{sarkar2000optimized} and fusion~\citep{manjikian1997fusion}.
In human development workflows, loop techniques are often avoided because they can reduce code readability and maintainability and are error-prone to implement correctly~\citep{mckinley1996improving, mens2004survey}.
These constraints may affect agentic workflows differently.
As software development becomes increasingly agent-centric, this gap presents an opportunity to investigate whether agents can apply optimization techniques that are traditionally underutilized by humans.

\subsection{From Presence of Validation to Completeness of Evidence}
Estimating performance impact is challenging, as sensitivity to workloads, hardware, and measurement noise often yields misleading or irreproducible results~\citep{mytkowicz2009producing, papadopoulos2020methodological}. 
Thus, rigorous improvements require validation through benchmarking or profiling~\citep{ beyer2019reliable, hoefler2015scientific}.
Our results show both progress and gaps in agentic validation practice.
Across the full sample, agentic PRs rely more often on static reasoning and human-authored PRs more often report benchmarks (Fig.~\ref{fig:validation_type_by_author}), but benchmark use among agentic PRs increases over time and reaches the human-authored rate by the end of the study period (Section~\ref{sec:result-rq3-temporal}).
This convergence is encouraging, yet using benchmarks does not establish that agents select evidence appropriate to a particular change.
In addition, agents may have learned to imitate the conventions of human benchmark reporting, including emitting plausible-looking results, and our data cannot distinguish improved measurement from improved presentation.

We also observe that human-authored PRs report validation evidence and quantitative metrics for code-smell refactorings less often than for resource-targeting performance changes, a pattern consistent with evidence reporting varying with the nature of the change; agentic PRs show no statistically significant corresponding difference (Section~\ref{sec:results-rq4-focused}).
In addition, a gap appears in trade-off reporting: caching and buffering PRs in both groups rarely document an adverse memory effect alongside a reported runtime improvement.
The assurance target should therefore be the appropriateness and completeness of evidence, not merely the presence of validation or a benchmark (Section~\ref{sec:results-rq4}).
Performance spans several metrics, and optimizations often exchange one resource for another (Section~\ref{sec:background-tradeoff}); even benchmark-based evidence can verify an intended gain while leaving relevant costs unobserved.
As the volume of agent-authored PRs grows, evaluation infrastructure should connect the optimization being attempted to representative workloads, expected metrics, and likely trade-offs.
Existing CI-based benchmarking, designed for human workflows, is unlikely to scale to agent-driven workloads, motivating standardized benchmarks and shared profiling platforms for automated validation.

\subsection{Trustworthiness under Increasing Autonomy}
As AI-assisted software engineering becomes more autonomous, human responsibility shifts toward specifying goals, constraints, and criteria for accepting a change. 
Model-Based Agentic Software Engineering (MAGE) frames this challenge in terms of \emph{Modeling}, which makes consequential knowledge and intent explicit, and \emph{Alignment}, which gives established obligations operational force~\citep{davis2026model}. 
For code optimization, a model could specify the target metric, workload, baseline, measurement protocol, and acceptable costs in other resources. Benchmark results can then be assessed against a stated objective rather than treated as sufficient evidence on their own.

Alignment would enforce the established parts of that model through benchmark and profiling measurements, statistical and resource checks, constraints, and acceptance gates. 
These checks matter because performance cannot be inferred from code alone: a functionally correct change may yield no improvement, regress on other workloads, or appear beneficial because of measurement noise or experimental conditions~\citep{mytkowicz2009producing, hoefler2015scientific, jain1990art, beyer2019reliable}. 
Passing tests also cannot rule out other defects in LLM- and agent-authored patches~\citep{xia2023automated, peng2026correct, sajadi2025safe}. 
In our sample, reported validation becomes more common over time (Section~\ref{sec:result-rq3-temporal}), but quantitative evidence and trade-off reporting remain limited (Section~\ref{sec:results-rq4}), and PR descriptions provide only authors' accounts of what was measured.
Although our artifacts do not reveal the projects' governed engineering environments, the results show why validation cannot be merely another agent activity: trustworthy acceptance must join an explicit performance model, independent evidence, and enforceable obligations, while retaining human authority over unresolved trade-offs and the sufficiency of that evidence.

\subsection{Implications for Longitudinal Agentic SE Datasets}
The temporal analysis in RQ3 (Section~\ref{sec:result-rq3-temporal}) shows that agent validation practices can change substantially within a single collection window.
An Agentic SE dataset therefore captures a time-bound state of practice rather than a stable characterization of agent behavior.
Conclusions about differences between agentic and human-authored PRs may consequently depend on the observation period, even when the same analysis is applied.

Reproducible longitudinal comparisons also require transparent authorship attribution.
Our conservative filtering (Section~\ref{sec:data-prep-current}) shows that the human-authored group is defined by the absence of selected automation and AI-assistance signals, rather than by verified human authorship.
Dataset releases should therefore preserve the raw attribution signals, record when they were collected, and version the classification rules applied to them.
Where possible, they should distinguish agentic, observably AI-assisted, and human-authored PRs.
Combined with time-balanced sampling or period-specific reporting, these practices would help future studies distinguish changes in development behavior from changes in attribution coverage or classification rules.

\section{Conclusion}

We extend our pilot study of 407 performance PRs to a larger sample of 2{,}260 agentic and human-authored PRs collected through June 2026. Using a 59-pattern catalog and evidence reported in PR artifacts, we compare adoption and patch characteristics and examine optimization strategies, validation practices, and reported performance metrics.
Our results show that agentic and human-authored PRs use broadly similar optimization strategies and report validation at comparable rates, but agentic PRs rely more on static reasoning and human-authored PRs more on benchmarks; agentic benchmark use increases over time.
Across both groups, only about half of PRs with validation evidence report an extractable performance metric, and relevant trade-offs are rarely documented.

These findings shift the central challenge from generating plausible optimizations to demonstrating that they work.
As agents assume more autonomous roles, performance changes should be accepted only when the evidence measures the intended effect under representative conditions and accounts for relevant resource trade-offs.
Evaluation pipelines must therefore make quantitative, multidimensional validation a requirement rather than treating any validation statement as sufficient.
More broadly, trustworthy agentic software engineering depends not only on what agents can produce, but also on engineering systems that make their objectives explicit, their evidence independently verifiable, and their acceptance conditions enforceable.

\section{Declarations}
\subsection{Funding}
This work was supported by NSF award \#2343596.
\subsection{Ethical Approval}
Not applicable. This study analyzes publicly available software repository data and does not involve human participants.
\subsection{Informed Consent}
Not applicable.
\subsection{Author Contributions}
Huiyun Peng contributed to idea formalization, data collection, methodology development, implementation, data analysis, and manuscript writing. Ricardo Calvo contributed to implementation, data collection, data analysis, and manuscript writing. Kelechi G. Kalu contributed to methodology development, data analysis, and manuscript writing. James C. Davis contributed to idea formalization and manuscript writing.
\subsection{Data Availability Statement}
Data are available on Hugging Face~\citep{perfpatchstudy2026dataset}; scripts and prompts are available in the replication repository~\citep{perfPatchStudy}. 

\subsection{Conflict of Interest}
Not applicable.

\bibliographystyle{spbasic}
\bibliography{references/reference}

\end{document}